\documentclass[pra,twocolumn,showpacs,superscriptaddress, floatfix, nofootinbib]{revtex4-1}
\usepackage[caption=false]{subfig}
\usepackage{amsmath,graphicx}
\usepackage{amsmath,amssymb,mathrsfs,esint}
\usepackage{multirow}
\usepackage{algorithm}
\usepackage[noend]{algpseudocode}
\usepackage{comment}
\usepackage{tabularx}
\usepackage{bm}
\usepackage[normalem]{ulem}
\usepackage[bookmarks=true,
   colorlinks=true,
   linkcolor=blue,
   urlcolor=blue,
   citecolor=blue,
   bookmarks=true,
   hyperindex=true
]{hyperref}
\usepackage{longtable}

\newcommand{\termCorrect}{correct}

\begin{document}

\title{Quantum-inspired optimization for wavelength assignment}

\author{Aleksey S. Boev}
\affiliation{Russian Quantum Center, Skolkovo, Moscow 143025, Russia}

\author{Sergey R. Usmanov}
\affiliation{Russian Quantum Center, Skolkovo, Moscow 143025, Russia}

\author{Alexander M. Semenov}
\affiliation{Russian Quantum Center, Skolkovo, Moscow 143025, Russia}

\author{Maria M. Ushakova}
\affiliation{Russian Quantum Center, Skolkovo, Moscow 143025, Russia}

\author{Gleb V. Salahov}
\affiliation{Russian Quantum Center, Skolkovo, Moscow 143025, Russia}

\author{Alena S. Mastiukova}
\affiliation{Russian Quantum Center, Skolkovo, Moscow 143025, Russia}

\author{Evgeniy O. Kiktenko}
\affiliation{Russian Quantum Center, Skolkovo, Moscow 143025, Russia}

\author{Aleksey K. Fedorov}
\affiliation{Russian Quantum Center, Skolkovo, Moscow 143025, Russia}

\begin{abstract}
Problems related to wavelength assignment (WA) in optical communications networks
involve allocating transmission wavelengths for known transmission paths between nodes that minimize a certain objective function, for example, the total number of wavelengths.
Playing a central role in modern telecommunications, this problem belongs to NP-complete class for a general case, so that obtaining optimal solutions for industry relevant cases is exponentially hard.
In this work, we propose and develop a quantum-inspired algorithm for solving the wavelength assignment problem.
We propose an advanced embedding procedure for this problem into the quadratic unconstrained binary optimization (QUBO) form 
having an improvement in the number of iterations with price-to-pay being a slight increase in the number of variables (``spins''). 
Then we compare a quantum-inspired technique for solving the corresponding QUBO form against classical heuristic and industrial combinatorial solvers.
The obtained numerical results indicate on an advantage of the quantum-inspired approach in a substantial number of test cases against the industrial combinatorial solver that works in the standard setting.
Our results pave the way to the use of quantum-inspired algorithms for practical problems in telecommunications and open a perspective for the further analysis of the employ of quantum computing devices.
\end{abstract}

\maketitle

\section{Introduction}

Optimization is a tool with applications across various technologies~\cite{Paschos2014}.
However, solving complex real-world optimization problems is computationally intensive even in the case of using advanced, specialzed hardware.
Quantum computers are widely believed to be useful for solving computationally difficult optimization problems beyond the capability of existing computing devices is to use quantum optimization~\cite{Farhi2000,Das2008,Lidar2018,Fedorov2022,Farhi2014}.
A general approach consists in encoding a cost function in a quantum Hamiltonian~\cite{Lucas2014}, so that its low-energy state is obtained starting from a generic initial state.
Among existing methods to achieve such dynamics, quantum annealing offers physical implementations of a non-trivial size~\cite{Amin2021}.
Quantum annealing is by now explored for analysis of various areas,
such as chemistry calculations~\cite{Leib2019,Chermoshentsev2021}, lattice protein folding~\cite{Aspuru-Guzik2012-2,Fingerhuth2018}, genome assembly~\cite{Fedorov2021,Sarkar2021},
solving polynomial systems of equations for engineering applications~\cite{Sota2019} and linear equations for regression~\cite{Sota2019},
portfolio optimization~\cite{Orus2019,Orus2020,Grant2021,Alexeev2022}, forecasting crashes~\cite{Orus2019-2}, 
finding optimal trading trajectories~\cite{Rosenberg2016-2}, optimal arbitrage opportunities~\cite{Rosenberg2016},  optimal feature selection in credit scoring~\cite{Rounds2017},
foreign exchange reserves management~\cite{Vesely2022},
traffic optimization~\cite{Neukart2017,Inoue2021,Hussain2020},
scheduling~\cite{Venturelli2016,Ikeda2019,Sadhu2020,Botter2020,Domino2021,Domino2021-2}, railway conflict management~\cite{Domino2021,Domino2021-2}, and many others~\cite{Fedorov2022}.
Advances also include the recent experimental demonstration of a superlinear quantum speedup in finding exact solutions for the hardest maximum independent set graphs~\cite{Lukin2022}.

Although quantum optimization algorithms suggest an intriguing possibility to solve computationally difficult problems beyond the capability of classical computers, 
exiting conceptual and technical limitations make it challenging to use it for solving problems of industry relevant sizes. 
Attempts to simulate quantum computations classically resulted in a new class of algorithms and techniques know as {\it quantum-inspired}~\cite{Tiunov2019,Lloyd2019}.
As soon as these algorithms are compatible with currently existing (classical) hardware, analyzing their limiting capabilities and advantages over classical approaches are required towards their use in practice. 
Specifically, a way to solve combinatorial optimization problems via simulating the coherent Ising machine (SimCIM) has been proposed~\cite{Tiunov2019}.
SimCIM algorithm is able to solve optimization problems that are formulated in the quadratic unconstrained binary optimization (QUBO) / Ising form, which can be done for various practically relevant cases~\cite{Lucas2014}.
The SimCIM approach has demonstrated capabilities to outperform {\it bona fide} coherent Ising machine and existing classical methods for certain GSet graphs.  
However, one of the arising questions is related to the need in to tune hyperaparameters~\cite{Tiunov2019}. 
For a wide range of benchmark of quantum-inspired heuristic solvers for quadratic unconstrained binary optimization, 
namely D-Wave Hybrid Solver Service, Toshiba Simulated Bifurcation Machine, Fujitsu Digital Annealer, and simulated annealing on a personal computer, see also Ref.~\cite{Oshiyama2022}.

Design of optical communication network is a specific industrial avenue, in which combinatorial optimisation in ubiquitous.
Examples of tasks include finding optimal transmission and reservation paths, frequency allocation, network throughput maximization and many others~\cite{Resende2003,Vesselinova2020}.
A notable example is the routing and wavelength assignment (RWA) problem, which consists in allocating transmission wavelengths and finding transmission paths between nodes that minimize the total number of wavelengths.
Conventional techniques, such as linear programming and mixed integer programming, are useful for most of the cases; however, 
the combinatorial nature and hardness of the problems make it extremely challenging to apply these techniques for large-scale problems.
It is then reasonable to assume that telecommunication industry may benefit from the use of quantum-inspired algorithm in the near-term horizon and quantum computing in the future~\cite{Martin2021,Harwood2021}.

In this study, we consider the variant of the RWA problem. 
To explain more precisely, we focus on the wavelength assignment task for known routes which we further refer to as the  wavelength assignment (WA) problem.
This problem is generally NP-hard, so its solution is computationally challenging for large sizes.
We propose an original way to transform the WA problem to the QUBO form, which makes it compatible with quantum-inspired optimization algorithm and, in principle, quantum annealing hardware.
For solving this problem, we develop a technique based on the SimCIM quantum-inspired optimization solver~\cite{Tiunov2019} with the use of the Lagrange multipliers for minimizing the number of hyperparameters.
Our numerical results indicate on an advantage of the quantum-inspired solver in a number of test cases against the industrial combinatorial solver working on the standard settings.

\section{Wavelength assignment problem (WA)}\label{sec:WA}

\begin{figure}[t]
	\vskip -3mm
	\includegraphics[width=\linewidth]{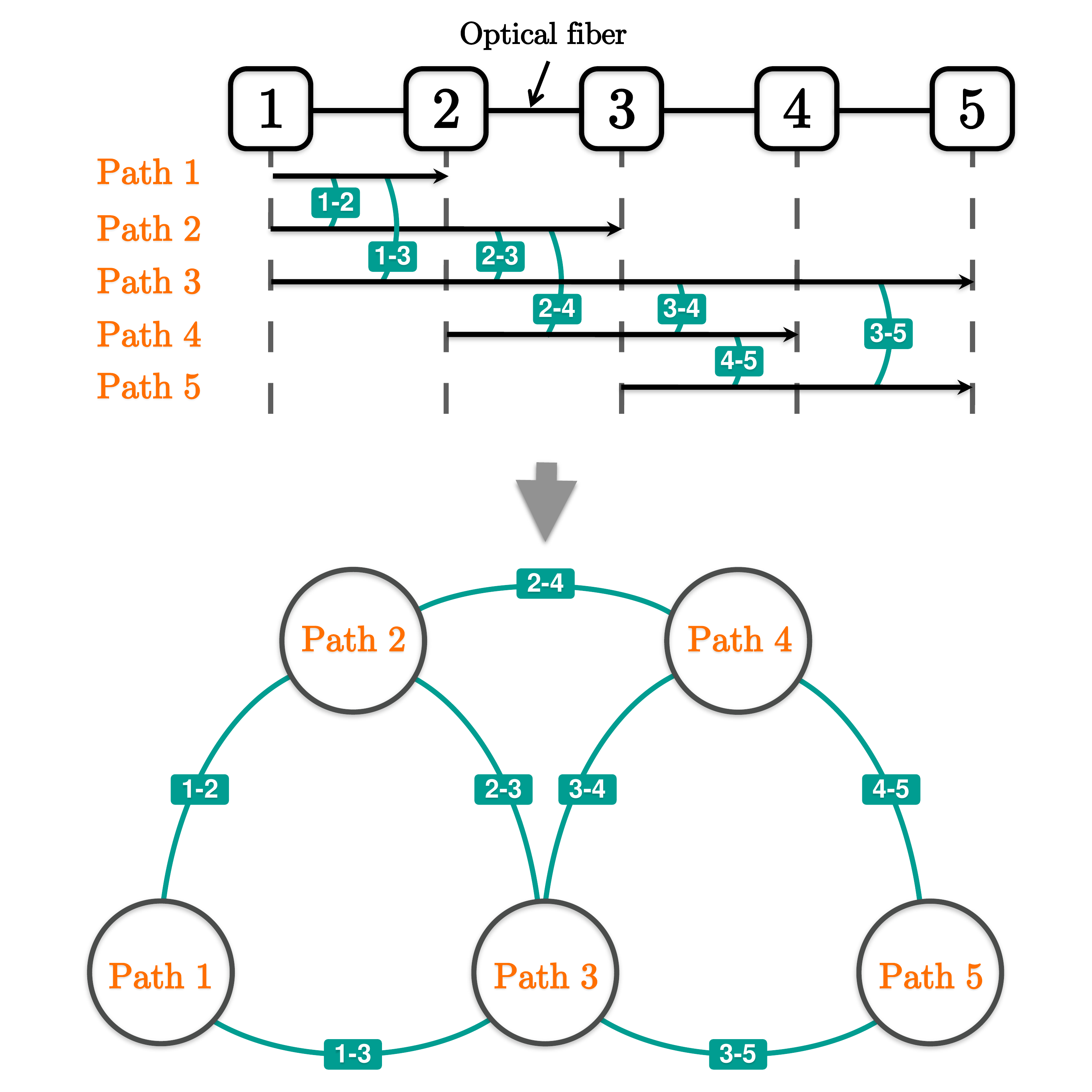}
	\caption{Illustration of the approach.
	A linear network with generated requests and paths consisting of 5 nodes, 4 edges, and 5 traffic paths is considered: Solid lines represent original edges, and the arrows lines represent traffic paths. 
	One can reduce the WA problem to a graph coloring problem with a simple graph transformation (bottom of the figure): 
	Each traffic path is now considered a vertex; if two traffic paths share (at least) one fiber they are connected by an edge.}
	\label{fig:rwa_example}
\end{figure}

\begin{figure}[t]
	\vskip -3mm
	\includegraphics[width=\linewidth]{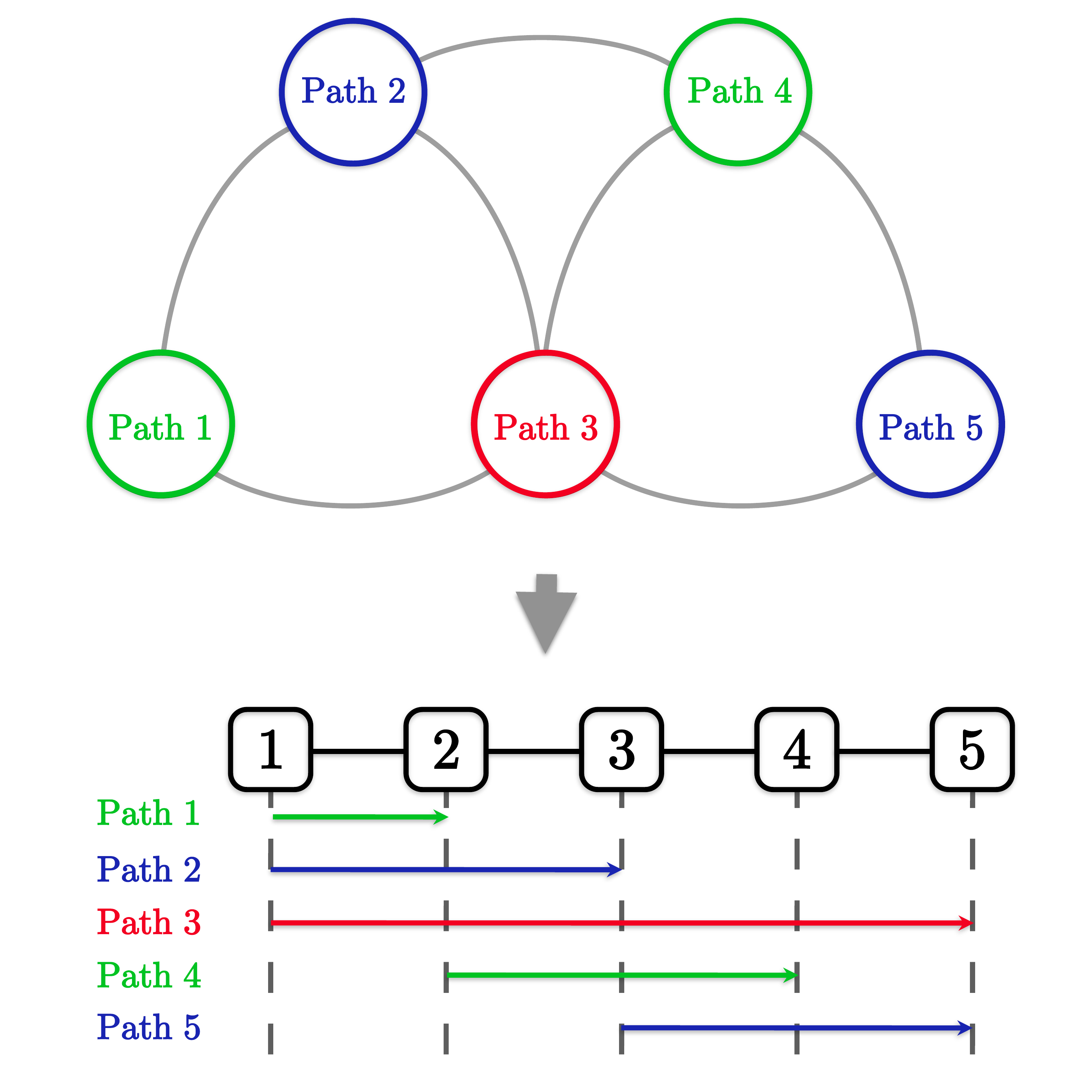}
	\caption{An example of a graph coloring problem and its representation to the network graph with requests.}
	\label{fig:rwa_convertion}
\end{figure}

Let us consider a network connecting a number of endpoints with optical links (see an example in Fig.~\ref{fig:rwa_example}.) 
Several endpoints that are interconnected by optical links sequentially comprise a path between transmitter and receiver. 
A single optical link can be shared between several paths given that each path is assigned different wavelengths.
Each path is indicated by the path ID, which uniquely identifies a pair of transmitting/receiving nodes, sequence of interconnecting nodes, and the wavelength ID.

The WA problem implies allocation of the wavelength IDs for paths that are pre-computed and known {\textit{a priori}} in such a way to meet the target objective, for example, the number of the used wavelengths is minimized
\footnote{We note that other objectives for optimization are also possible, such as total throughput or network resiliency.}. 
Formally, WA is considered to be \termCorrect{} if and only if it satisfies the following requirements:
(i) each path should use a single wavelength and 
(ii) several paths sharing the same edge should have different wavelengths.

The problem of finding \termCorrect{} wavelength allocation under given constraints is equivalent to the coloring problem~\cite{Lucas2014} in transformed graph $G = (V,E)$, 
where nodes $V$ and edges $E$ representing paths and their intersections in fibers, correspondingly (two nodes from $V$ are connected if and only if the corresponding paths have an intersection within the optical network).
Let $N_V$ and $N_E$ denote numbers of vertices and edges of $G$, respectively.
Later we interchangeably use terms {\it wavelengths} and {\it colors} since the underlying problems are formally identical. 
The example of the correspondence of network paths to graph coloring mapping is shown in Fig.~\ref{fig:rwa_convertion}.

In order to define a particular coloring of graph $G$ with at most  $W$ colors, we introduce a two collections of auxiliary variables.
The first variable is $\boldsymbol x$ that consists of $N_V W$ binary variables 
\begin{align}
	x_{vi} &=
	\begin{cases}
		1, & \text{if vertex $v$ is assigned wavelength $i$}, \\
		0, & \text{otherwise}.
	\end{cases}
\end{align}
The second one, denoted by  $\boldsymbol w$, consists of $W$ binary variables
\begin{align}
	w_{i} &=
	\begin{cases}
		1, & \text{if $i$-th wavelength is assigned}, \\
		0, & \text{otherwise}.
	\end{cases}
\end{align}

Employing $\boldsymbol x$ and $\boldsymbol w$, the problem of finding an \termCorrect{} allocation with minimum number of the used wavelengths not exceeding some maximal number $W\geq 1$, 
can be formulated as an integer programming (IP) problem of the following form:
\begin{align}
	&\sum_{i=1}^{W} w_{i} \rightarrow \min, \quad \text{s.t.} \label{cnst:0}\\  
	&\sum_{i=1}^{W} x_{vi} = 1 \quad \forall v \in V, \label{cnst:1}\\
	&x_{u i} + x_{v i} \leq w_{i} \quad \forall i \in\{1, \ldots, W\},  \forall (u, v) \in E. \label{cnst:2}
\end{align}
One can see that constraint~(\ref{cnst:1}) assures that each vertex is assigned to exactly one wavelength, while constraint~(\ref{cnst:2}) indicates that two adjacent vertices are not assigned the same wavelength.

This problem is generally NP-hard, so its solution is computationally challenging for large sizes. 
As it is shown below, the QUBO reduction makes the problem compatible with quantum-inspired algorithms that can shift tractability boundaries to higher problem sizes. 
While such reduction usually involves additional overheads in the problem size due to auxiliary variables, the overheads can be compensated by the computational advantage of quantum-inspired solvers leading to better overall results. 

\section{Results}

\subsection{Transforming the WA problem to a QUBO form}

In order to make the WA problem compatible with the SimCIM quantum-inspired optimization algorithm~\cite{Tiunov2019}, 
we first consider a transformation, allowing one to convert the IP problem~\eqref{cnst:0}--\eqref{cnst:2} into a QUBO form as follows:
\begin{equation} \label{eq:QUBO-form}
	\boldsymbol s^T  Q \boldsymbol s \rightarrow \min
\end{equation}
for a certain binary vector $\boldsymbol s$ and the symmetric real matrix $Q$.
This problem is equivalent to finding a configuration of binary-state particles (``spins'') that minimizes the energy
\begin{equation}
	{\cal H}(\boldsymbol s) = \boldsymbol s^T  Q \boldsymbol s,
\end{equation}
where the Ising Hamiltonian ${\cal H}$ consists of only single-order terms (energies of individual spins in external magnetic field) and pair-wise interactions between spins. 
Although spin variables usually are considered to take values $\pm1$, the transition to a binary form is quite straightforward~\cite{Fedorov2021}.

A known way~\cite{Lucas2014} to transform a graph coloring problem to the QUBO form, is to set $\boldsymbol s:=\boldsymbol x$ (here we treat $\boldsymbol x$ as a $N_VW$-dimensional vector), and use the Hamiltonian of the form
\begin{equation}\label{hamiltonian:lucas}
    {\cal H}(\boldsymbol x)= {\cal H}_1(\boldsymbol x) + {\cal H}_2(\boldsymbol x), 
\end{equation}
where
\begin{align}
	\mathcal{H}_1(\boldsymbol x) &= \sum_{v=1}^{N_V}\left(1-\sum_{i=1}^{W}x_{v i}\right)^{2},\\
	\mathcal{H}_2(\boldsymbol x) &= \sum_{(u,v)\in E}^{}\sum_{i=1}^{W} x_{ui}x_{vi}.
\end{align}

One can see that $\mathcal{H}_1(\boldsymbol x)>0$ in the case where single node is assigned with two distinct colors, while $\mathcal{H}_2(\boldsymbol x)>0$ when two adjacent vertices are assigned the same color. 
If minimization routine provides some $\boldsymbol x$ such that ${\cal H}(\boldsymbol x)=0$, then $\boldsymbol x$ defines a \termCorrect{} coloring with at most $W$ colors.
Therefore, an ability to solve the QUBO problem corresponding to Hamiltonian~\eqref{hamiltonian:lucas} garantees one to solve a decision problem of whether it is possible to color a graph with at most $W$ colors.
Since it is always possible to color a graph with $W=N_V$ colors, a minimal number of colors can be obtained, for example, by using a standard binary search with at most $\lceil\log_2(N_V)\rceil$ iterations.
We note that this approach is quite sensitive to possible imperfections of QUBO problem solutions, especially at first iterations of the binary search.
An alternative way is to decrease $W$ by unit at each step, that however, results in a possible increase of iteration numbers up to ${\cal O}(W_{\rm start})$, where $W_{\rm start}$ is the initial upper bound for colors number.

\subsection{Improving QUBO transformation for quantum-inspired annealing}

We propose an improved approach for solving graph coloring problem by developing an alternative transformation into a QUBO form. 
In our approach we pursue two major goals.
The first is decreasing the number of QUBO problems to be solved.
The second is making the whole algorithm robust against the possibility of finding not optimal, but some suboptimal solution for a particular QUBO problem.
We note that these points are of particular importance in the framework of using (quantum-inspired) annealing for solving QUBO problems.

The main idea of our approach is to consider an extended $N_V(W+1)$-dimensional binary vector $\boldsymbol s:=(\boldsymbol w, \boldsymbol x)$ and take the target Hamiltonian in the following form:
\begin{multline}\label{hamiltonian:full}
	\mathcal{H} (\boldsymbol w, \boldsymbol x) = c_0\mathcal{H}_0(\boldsymbol w) + c_1\left[\mathcal{H}_1(\boldsymbol x) + \mathcal{H}_2(\boldsymbol x)\right] \\+ c_2\mathcal{H}_3(\boldsymbol w, \boldsymbol x),
\end{multline}
where
\begin{align}
	\mathcal{H}_0(\boldsymbol w) &= \sum_{i=1}^{W}w_i, \\
	\mathcal{H}_3(\boldsymbol w, \boldsymbol x) &= \sum_{(u,v)\in E}^{}\sum_{i=1}^{W}(1-w_i)\left(x_{ui}+x_{vi}\right),    
\end{align}
and $c_i$ are positive coefficients satisfying a particular constraint (see more details in Methods~\ref{app:Hamiltonian}).
Minimization of this Hamitonian provides us the solution vector $(\boldsymbol w, \boldsymbol x)$ such that the optimal number of wavelength is encoded in $\boldsymbol w$ by non-zero values.
We note that the term $\mathcal{H}_0(\boldsymbol w)$ grows with the total number of used wavelengths;
$\mathcal{H}_1(\boldsymbol x)$ and $\mathcal{H}_2(\boldsymbol x)$ have the same form as in Eq.~(\ref{hamiltonian:lucas}); and $\mathcal{H}_3(\boldsymbol w, \boldsymbol x)$ 
is responsible for the relationship $w_i\geq x_{vi}$, which becomes positive when the relation is violated. 
Both terms $\mathcal{H}_2(\boldsymbol x)$ and $\mathcal{H}_3(\boldsymbol w, \boldsymbol x)$ correspond to inequalities (\ref{cnst:2}) in the IP form (see Methods~\ref{sec:methods}).

The complete algorithm of solving graph coloring problem (WA problem) is shown in Algorithm~\ref{alg:fast}.
The algorithm employs a subroutine ${\sf make\_qubo}(G,W)$ that generates the corresponding QUBO matrix $Q$ with respect to Hamiltonian~\eqref{hamiltonian:full}, given input graph $G$ and the target number of the wavelengths $W$.
The QUBO problem is then solved with subroutine ${\sf solve\_qubo}(Q)$, 
which finds the optimal spins vector $\boldsymbol s=(\boldsymbol w, \boldsymbol x)$ using the quantum-inspired SimCIM approach for the QUBO matrix $Q$ as defined in Ref.~\cite{Tiunov2019}.
In order check the validness of obtained solution, we use ${\sf check\_coloring}(G, \boldsymbol x)$ that validates the fulfilment of Eq.~\eqref{cnst:1} and Eq.~\eqref{cnst:2}.

\begin{algorithm}[H]
\caption{
Solving graph coloring problem with improved transformation
}\label{alg:fast}
\begin{algorithmic}[1]
\Require{$W$ is the initial upper bound on the number of wavelengths}
\Require{${\sf make\_qubo}(G, W)\,\to\, Q$}
\Require{${\sf solve\_qubo}(Q) \,\to\, (\boldsymbol w, \boldsymbol x)$}
\Require{${\sf check\_coloring}(G, \boldsymbol x)\,\to\, \rm true/false$}
\Statex{returns true if coloring is \termCorrect{}}
\newline

\State{$\boldsymbol x^{\rm opt} := \boldsymbol 0$ \Comment initializing solution variable}
\State $W' := W$ \Comment current number of colors
\While{$W' \geq 1$}
  \State{$Q:= {\sf make\_qubo}(G, W')$}
  \State{$(\boldsymbol w, \boldsymbol x):={\sf solve\_qubo}(Q)$}
\If {${\sf check\_coloring}(G, \boldsymbol x)={\rm true}$}
  \State{$W' := \sum_{i=1}^{W'} w_i-1$}
  \State{$\boldsymbol x^{\rm opt} := \boldsymbol x$ }
  \Else
    \State {break}
  \EndIf
\EndWhile

\State \Return $\boldsymbol x^{\rm opt}$

\end{algorithmic}
\end{algorithm}

One can see that, if ${\sf solve\_qubo}(Q)$ provides an optimal solution, then the whole problem is solved in the first iteration.
However, even in the case when the obtained solution is sub-optimal, the updated problem with the reduced upper bound $W$ becomes easier to solve, and the algorithm converges with a few numbers of iterations.

\subsection{Numerical results}\label{sec:bench}

Here we solve the WA problem and obtain results with use of (i) the proposed technique based on quantum-inspired optimization SimCIM~\cite{Tiunov2019} (with the improved approach, see Methods), 
(ii) industry grade commercial Gurobi optimization software, and (iii) open-source mixed integer programming solver --- GLPK. 
We note that in the case of the quantum-inspired optimization with SimCIM, we solve the problem in the QUBO form~\eqref{hamiltonian:full}, whereas in case of Gurobi and GLPK we use the IP formulation of graph coloring [see Eqs.~(\ref{cnst:0})--(\ref{cnst:2})].
Additionally, we include the results obtained via largest degree first (LDF) heuristics used as the baseline, since it allows one to instantly produce feasible coloring without numerical optimization. 
We also ran the experiments for original QUBO transormation proposed in Ref.~\cite{Lucas2014} and compared them to our proposed QUBO in the Table~\ref{tab:qubocomparision} in Appendix.

\begin{table}[h]
\centering
\resizebox{0.8\columnwidth}{!}{%
\begin{tabular}{|c|c|c|c|c|}
\hline
\textbf{\begin{tabular}[c]{@{}c@{}}Number \\ of nodes\end{tabular}} & \textbf{LDF} & \textbf{GLPK} & \textbf{Gurobi} & \textbf{SimCIM} \\ \hline
\textbf{10}  & 4.46  & 4.34 & 4.34  & 4.34  \\ \hline
\textbf{20}  & 6.82  & 6.36 & 6.36  & 6.36  \\ \hline
\textbf{30}  & 9.03  & 8.03 & 8.02  & 8.02  \\ \hline
\textbf{40}  & 10.92 & -    & \textit{\textbf{9.38}}  & 9.39  \\ \hline
\textbf{50}  & 12.80 & -    & \textit{\textbf{10.88}} & 10.96 \\ \hline
\textbf{60}  & 14.83 & -    & \textit{\textbf{12.28}} & 12.44 \\ \hline
\textbf{70}  & 16.62 & -    & \textit{\textbf{13.70}} & 14.01 \\ \hline
\textbf{80}  & 18.41 & -    & \textit{\textbf{15.34}} & 15.56 \\ \hline
\textbf{90}  & 20.10 & -    & 17.21 & \textit{\textbf{17.02}} \\ \hline
\textbf{100} & 22.01 & -    & 19.64 & \textit{\textbf{18.54}} \\ \hline
\end{tabular}%
}
\\ \textbf{Average number of colors \\ (lower is better)}
\caption{Numerical results obtained with largest degree first (LDF) heuristics, open-source mixed integer programming solver (GLPK),
Gurobi optimization software, and SimCIM quantum-inspired optimization 
on number of colors averaged by number of nodes. The best result is highlighted in bold.}
\label{tab:benchmark_nc}
\end{table}

Our numerical experiments have been performed on a synthetic dataset of 900 randomly generated graphs with varying nodes number and edge probability (for details, see Methods~\ref{app:dataset}). 
The main characteristics that we are interested in are time-to-solution (TTS) and the number of colors in the obtained solution.
The total runtime has been limited by 300 seconds, and the the best solutions have been compared. 
Results are averaged over 90 runs for each graph size (for details, see Table~\ref{tab:benchmark_nc}).
For all numerical experiments, we use the same hardware set, which is based on Xeon E-2288G 3,7GHz CPU, 128GB RAM, and GeForce GTX1080 8GB graphics card.

\begin{table}[H]
\vskip 3mm
\centering
\resizebox{0.7\columnwidth}{!}{%
\begin{tabular}{|c|c|c|c|}
\hline
\textbf{\begin{tabular}[c]{@{}c@{}}Number \\ of nodes\end{tabular}} & \textbf{GLPK} & \textbf{Gurobi} & \textbf{SimCIM} \\ \hline
\textbf{10}  & 1.77   & 0.002  & 0.19   \\ \hline
\textbf{20}  & 103.97 & 0.02   & 0.45   \\ \hline
\textbf{30}  & 195.39 & 0.12   & 4.95   \\ \hline
\textbf{40}  & -      & \textit{\textbf{0.79}}   & 8.90   \\ \hline
\textbf{50}  & -      & \textit{\textbf{14.63}}  & 16.82  \\ \hline
\textbf{60}  & -      & 38.89  & 28.51*  \\ \hline
\textbf{70}  & -      & 66.01  & 61.58*  \\ \hline
\textbf{80}  & -      & 102.14 & 69.00*  \\ \hline
\textbf{90}  & -      & 144.23 & \textit{\textbf{79.87}}  \\ \hline
\textbf{100} & -      & 127.33 & \textit{\textbf{123.13}} \\ \hline
\end{tabular}%
}
\\ \textbf{Average time (seconds) \\ (lower is better)}
\caption{Mean solution time depending on the number of nodes for GLPK, Gurobi and SimCIM. The best result is highlighted in bold.
\\ * cases, where average number of colors is higher}
\label{tab:benchmark_tts}
\end{table}

Our results indicate that the quantum-inspired technique SimCIM demonstrates behaviour comparable with Gurobi in the case of small (10-30 nodes).
Moreover, the runtime of SimCIM is better for large-scale (90 and 100 nodes) graphs as it is indicated in Table \ref{tab:benchmark_tts}. 
Such trend can be explained. As the number of nodes increases, the number of inequalities in the ILP formulation of the problem grows rapidly. The number of inequalities is equal to the product of the number of edges by the number of colors available for coloring the vertices of the graph. So the complexity of the problem for the ILP solver increases rapidly with the number of nodes.
GLPK shows a stable result up to 30 nodes and becomes unstable further after a timeout interrupt without any solution with more than 10 percent instances.
We note that the comparison between our quantum-inspired approach and Gurobi is conducted in the common setting, so its additional tuning for obtaining better results is also possible.
At the same time, we find it interesting that quantum-inspired technique shows comparable or superior results in harder, industry relevant combinatorial optimization problem.

\subsection{Other potential applications}

While our goal was to demonstrate the applicability of quantum-inspired graph coloring algorithm for wavelength assignment problem, our approach can be applied to a variety of problems, in particular from the field of scheduling~\cite{marx2004graph}.

Assuming we have the set of jobs to schedule, every job requires one time slot and some jobs can not be executed at the same time due to some interference with each other, we need to determine the minimal time when every job will be finished or how many time slots they will occupy. One can build the graph, so that vertices correspond to the jobs and two vertices are connected if these jobs can't be executed at the same time. The colors of vertices represent time slots to assign, so graph has k number of colors if the jobs can be executed in k time slots.

Using our approach we take proposed Hamiltonian in Eq.~(\ref{hamiltonian:full}) and redefine its variables so that 
\begin{align}
	x_{vi} &=
	\begin{cases}
		1, & \text{if vertex $v$ is assigned time slot $i$}, \\
		0, & \text{otherwise},
	\end{cases}
\end{align}
and
\begin{align}
	w_{i} &=
	\begin{cases}
		1, & \text{if $i$-th time slot is assigned}, \\
		0, & \text{otherwise}.
	\end{cases}
\end{align}

That way the jobs scheduling problem can be solved using quantum-inspired annealing analogously to WA problem.  

The same approach can be implemented for tasks from other fields, such as computer register allocation~\cite{chaitin1981register}, storage of chemicals~\cite{ott2020chemstor} and printed circuit board testing~\cite{garey1976application}.

\section{Conclusion}\label{sec:conclusion}

A search for new approaches to solving practically-relevant optimization problems is a clear goal for many industrial applications since even minor improvement on a large scale may generate serious economical impact.
In this domain, much attention is paid to quantum computing, which is believed to be useful for such class of problems.
At the current technological level, practical quantum advantage, for example, in optimization is still needed to be achieved. 
An interesting part of this research is understanding of the physical origin of the potential advantages of quantum computing technologies. 
Attempts to simulate quantum computation classically resulted in a new class of algorithms and methods know as quantum-inspired, which are ready to be tested for industry-relevant problems. 

In this work, we have considered the industry important problem in the field of telecommunications. 
We have demonstrated a way to make it compatible with quantum and quantum-inspired techniques. 
Interestingly, our numerical results have indicated on an advantage of the quantum-inspired solver in a number of test cases against the industrial combinatorial solver working on the standard settings.

One may expect that the additional tuning of the industry grade commercial optimization solver may result in a substantial improving of its performance. 
At the same time, studying the origins of the advantages of the quantum-inspired approach, which are largely beyond the scope of the present proof-of-concept demonstration, would allow its further progress as well. 

We would like to note that our comparison is limited by the upper bound of 100 nodes, since it allows us to run all solvers in equivalent hardware setup using CPU mode on single core. Further analysis of larger graphs requires running SimCIM solver on GPU card, which gives significant acceleration factor not directly available in conventional MIP algorithms, which are heavily dependent on graph processing routines. 
As for the multi-core CPU execution environment, some MIP solvers can benefit from such setup by running various optimization strategies and hyperparameters simultaneously. 
Such speed up quickly reaches saturation point at the level of 8~16 cores (with around ~2x improvement in accordance with Gurobi experiments, see slide 26~\cite{glockner2015parallel}) and demonstrates no substantial improvement at higher concurrency levels. 
On the other side, quantum-inspired approach exploits parallelism on the level of starting optimization points, which demonstrates slower, but stable performance increase at the higher levels of concurrency (100 $\sim$ 1000 parallel units of execution). 
Thus, we conduct our benchmarks exclusively using CPU mode on single-core to avoid bias towards either solution approach. In order to maintain fairness of comparison for larger graphs our benchmark routine should be further revised to account for heterogeneous (CPU/CPU multi-core vs GPU/multi-GPU) computing environments.

\section*{Acknowledgements}

We acknowledge use of the Gurobi for this work; the views expressed are those of the authors and do not reflect the official policy or position of Gurobi. 
The part of this work related to the analysis of quantum-inspired optimization algorithm was supported by Russian Science Foundation (19-71-10092).

\bibliographystyle{naturemag} 
\bibliography{bibliography.bib}

\section{Methods}\label{sec:methods}

\subsection{Hamiltonian of wavelength assignment problem}\label{app:Hamiltonian}
    
The main step in solving an optimization problem using quantum and quantum-inspired annealing is to map the problem of interest to the energy Hamiltonian (so-called Ising Hamiltonian), 
so the quantum device could find the ground state that corresponds to the optimum value of the objective function. 
Here we formulate a mapping of the graph coloring problem into QUBO form given by Eq.~\eqref{eq:QUBO-form}. 
There is a well-known transformation of the graph $G=(V,E)$ coloring decision model~\cite{Lucas2014} that shows possibility of coloring with some constant number of colors $W$, 
but we represent novel QUBO transformation that could minimize number of colors and implement original problem statement Eq.~(\ref{cnst:0}-\ref{cnst:2}). 

The objective function $\sum_{i=1}^W w_i$ could be exactly mapped to the QUBO form:
\begin{equation}\label{qubo:0}
	\mathcal{H}_0(\boldsymbol w) = \sum_{i=1}^{W}w_i,
\end{equation}
where, recall, $\boldsymbol w=(w_1,\ldots, w_W)$ is a binary vector indicating colors used in coloring.
The constraint $\sum_{i=1}^W x_{vi}=1$ for every $v\in V$ after mapping takes the form
\begin{equation} \label{qubo:1}
	\mathcal{H}_1(\boldsymbol x) = \sum_{v=1}^{N_V}\left(1-\sum_{i=1}^{W}x_{v i}\right)^{2},
\end{equation}
where $N_V$ is the number of nodes in $G$.

The situation with the second constraint $x_{u i} + x_{v i} \leq w_{i}$ for every $i \in\{1, \ldots, W\}$ and $(u, v) \in E$ appears to be more complicated.
One can see that it involves three variables, and thus can not be directly embedded into a two-body Hamiltonian.
However, we can use the following trick.
One can easily  check that for arbitrary $a,b,c\in\{0,1\}$, the following equivalence holds:
\begin{equation}\label{cnst:2_abc}
    a+b\leq c \Leftrightarrow 	
    	\begin{cases}
		ab=0, \\
		(1-c)(a+b)=0.
	\end{cases}
\end{equation}
This fact allows us to embed the conditions  $x_{u i} + x_{v i} \leq w_{i}$  into two Hamiltonians:
\begin{eqnarray} 
	\mathcal{H}_2(\boldsymbol x) &=& \sum_{(u,v)\in E}^{}\sum_{i=1}^{W} x_{ui}x_{vi}, \\
	\mathcal{H}_3(\boldsymbol w, \boldsymbol x) &=& \sum_{(u,v)\in E}^{}\sum_{i=1}^{W}(1-w_i)\left(x_{ui}+x_{vi}\right). 
\end{eqnarray}

The resulting Hamiltonian consist of all components sum:
\begin{multline}\label{hamiltonian:full_}
	\mathcal{H}(\boldsymbol w, \boldsymbol x) = c_0\mathcal{H}_0(\boldsymbol w) + c_1\left[\mathcal{H}_1(\boldsymbol x) + \mathcal{H}_2(\boldsymbol x)\right]
	 \\ + c_2\mathcal{H}_3(\boldsymbol w, \boldsymbol x),   
\end{multline}
where $c_0$, $c_1$, and $c_2$ are positive constants stand for a positive penalty value. 
We note that the sum $\mathcal{H}_1(\boldsymbol x) + \mathcal{H}_2(\boldsymbol x)$ is exactly matched with the classical decision problem~\cite{Lucas2014} and responsible for the correct coloring of the graph. 
Therefore, $\mathcal{H}_1(\boldsymbol x)$, $\mathcal{H}_2(\boldsymbol x)$ are grouped with the same penalty coefficient $c_1$.
Coefficients $c_0$, $c_1$, and $c_2$ should be set manually, using the following criteria: the penalty value $c_1$ should be high enough to keep the final solution from violating constraints. 
At the same time, too big penalty value can overwhelm the target function, making it difficult to distinguish solutions of different qualities. 
We establish inequalities for constraint coefficients that show the equivalence of IP and QUBO models of a problem.

{\bf Proposition (QUBO penalty coefficients selection).}
Consider an IP problem given by Eq.~\eqref{cnst:0}-\eqref{cnst:2} for some maximal colors number $W$ and some graph $G=(V,E)$ with $N_E$ edges.
If the IP problem has a solution, then the corresponding QUBO problem, given by Hamiltonian~\eqref{hamiltonian:full_} with penalty coefficients satisfying
\begin{eqnarray}
	c_1 &>& 2N_EWc_2+Wc_0,  \label{eq:c_cond_1}\\
	c_2 &>& Wc_0, \label{eq:c_cond_2}
\end{eqnarray}
has a solution, equivalent to the solution of the IP problem.

{\bf Proof.}
First, let us rewrite Hamiltonian \eqref{hamiltonian:full_} in the form
\begin{equation}
	{\cal H}(\boldsymbol w, \boldsymbol x) = c_0{\cal A}(\boldsymbol w) + c_1{\cal B}(\boldsymbol x) + c_2{\cal C}(\boldsymbol w, \boldsymbol x),
\end{equation}
where 
\begin{equation}
	\begin{aligned}
		{\cal A}(\boldsymbol w)&:=\mathcal{H}_0(\boldsymbol w),\\
		{\cal B}(\boldsymbol x)&:=\mathcal{H}_1(\boldsymbol x)+\mathcal{H}_2(\boldsymbol x), \\
		{\cal C}(\boldsymbol w, \boldsymbol x)&:=\mathcal{H}_3(\boldsymbol w, \boldsymbol x).
	\end{aligned}
\end{equation}
Note that ${\cal A}$, ${\cal B}$, and ${\cal C}$ can take non negative integer values only.
Let $(\boldsymbol w_{\rm I}, \boldsymbol x_{\rm I})$ and $(\boldsymbol w_{\rm Q}, \boldsymbol x_{\rm Q})$ be solutions of the IP and QUBO problems correspondingly.
Our goal is to prove that (i)
\begin{equation}
	{\cal B}(\boldsymbol x_{\rm Q})= {\cal C}(\boldsymbol w_{\rm Q}, \boldsymbol x_{\rm Q})=0,
\end{equation}
i.e., $(\boldsymbol x_{\rm Q},\boldsymbol w_{\rm Q})$ defines a correct coloring, and (ii)
\begin{equation}
	{\cal A}(\boldsymbol w_{\rm Q})=\sum_{i=1}^W  (\boldsymbol w_{{\rm I}})_i,
\end{equation}
i.e., the solution of the QUBO problem coincides with the one of the IP problem.

First, let us see that Eq.~\eqref{eq:c_cond_1} assures ${\cal B}(\boldsymbol x_{\rm Q})=0$.
The proof of this part is by a contradiction. 
Suppose that ${\cal B}(\boldsymbol x_{\rm Q}) \geq 1$.
Consider  the difference of energy functions
\begin{multline} \label{eq:eqs_list}
	\Delta {\cal H} :=
	{\cal H}(\boldsymbol w_{\rm Q}, \boldsymbol x_{\rm Q})-{\cal H}(\boldsymbol w_{\rm I}, \boldsymbol x_{\rm I}) \\
	=c_0 \left[ {\cal A}(\boldsymbol w_{\rm Q}) - {\cal A}(\boldsymbol w_{\rm I})\right]
	+c_1 \left[ {\cal B}(\boldsymbol x_{\rm Q}) - {\cal B}(\boldsymbol x_{\rm I})\right] \\
	+c_2 \left[ {\cal C}(\boldsymbol w_{\rm Q}, \boldsymbol x_{\rm Q}) - {\cal C}(\boldsymbol w_{\rm I}, \boldsymbol x_{\rm I})\right]. 
\end{multline}
The correctness of the IP solution implies ${\cal B}(\boldsymbol x_{\rm I})=0$, and so
$ {\cal B}(\boldsymbol x_{\rm Q}) - {\cal B}(\boldsymbol x_{\rm I})\geq 1$.
The differences in terms with ${\cal A}$ and ${\cal C}$ can be lower bounded by the corresponding extreme values:
\begin{eqnarray}
	&&{\cal A}(\boldsymbol w_{\rm Q}) - {\cal A}(\boldsymbol w_{\rm I}) \geq -W, \\
	&&{\cal C}(\boldsymbol w_{\rm Q}, \boldsymbol x_{\rm Q}) - {\cal C}(\boldsymbol w_{\rm I}, \boldsymbol x_{\rm I}) \geq -2 N_E W.
\end{eqnarray}
In this way, Eq.~\eqref{eq:eqs_list} transforms into
\begin{equation}
	\Delta {\cal H} \geq -c_0W + c_1 - 2 c_2  N_E W > 0,
\end{equation}
given constraint~\eqref{eq:c_cond_1}.
However, this result contradicts with the fact that $(\boldsymbol w_{\rm Q}, \boldsymbol x_{\rm Q})$ provides the minimal energy.
Therefore, ${\cal B}(\boldsymbol x_{\rm Q})=0$, and
\begin{equation}
	{\cal H}(\boldsymbol w_{\rm Q}, \boldsymbol x_{\rm Q})=c_0{\cal A}(\boldsymbol w_{\rm Q})+c_2 {\cal C}(\boldsymbol w_{\rm Q}, \boldsymbol x_{\rm Q}).
\end{equation}

We then prove that ${\cal C}(\boldsymbol w_{\rm Q}, \boldsymbol x_{\rm Q})$ is zero as well.
Indeed, if ${\cal C}(\boldsymbol w_{\rm Q}, \boldsymbol x_{\rm Q})\geq 1$ , then
\begin{multline}
	\Delta {\cal H} = c_0  \left[ {\cal A}(\boldsymbol w_{\rm Q}) - {\cal A}(\boldsymbol w_{\rm I})\right] \\+
	c_2 \left[ {\cal C}(\boldsymbol w_{\rm Q}, \boldsymbol x_{\rm Q}) - {\cal C}(\boldsymbol w_{\rm I}, \boldsymbol x_{\rm I})\right] \\\geq -c_0W + c_2 > 0,
\end{multline}
provided ${\cal C}(\boldsymbol w_{\rm I}, \boldsymbol x_{\rm I})=0$ and the second constraint~\eqref{eq:c_cond_2}.
Thus, ${\cal H}(\boldsymbol w_{\rm Q}, \boldsymbol x_{\rm Q})=c_0{\cal A}(\boldsymbol w_{\rm Q})$.

Finally, ${\cal A}(\boldsymbol w_{\rm Q})={\cal A}(\boldsymbol w_{\rm I})$, since otherwise, 
either there exist a solution for the QUBO problem that is better than  $( \boldsymbol w_{\rm Q}, \boldsymbol x_{\rm Q})$, or $(\boldsymbol x_{\rm I},\boldsymbol w_{\rm I})$  is not the true solution the IP problem.

Therefore, the optimal solution to the QUBO problem appears to be equivalent to the optimal solution to the corresponding IP problem.

\subsection{Wavelength assignment QUBO transformation}\label{app:QUBO}

Here we demonstrate how to construct an operator matrix $Q$ of our QUBO model for the WA problem.  
Recall that we take the binary vector of the QUBO problem in the form $\boldsymbol s=(\boldsymbol w, \boldsymbol x)$, i.e. enumerate $K = (N_V+1)W$ binary variables $s_{k}$ and link them to our model variables as follows:
\begin{align}\label{eq:notation_qubo}
	s_{k} &=
	\begin{cases}
		w_{k}, & \text{$k = 1,\dots, W$}, \\
		x_{u i}, & \text{$k = uW + i,$}
	\end{cases} 
\end{align}
where $u=1,\dots, N_V, i=1,\dots , W$.

The goal is to find vector $\boldsymbol s$ that minimizes quadratic form $\boldsymbol s^TQ \boldsymbol s$ and we show that it is equivalent to minimizing energy of Hamiltonian~(\ref{hamiltonian:full}). 
Let us denote $A$ the adjacency matrix of network graph $G=(V,E)$ so that $a_{uv}=1$ if $(u,v) \in E$ and $a_{uv}=0$ otherwise.
We note that sum of $v$-th column of $A$ equals the degree of the vertex $v$ and the sum of all vertex degrees is $2N_E$. 
We rewrite operator (\ref{hamiltonian:full}) terms $\mathcal{H}_0(\boldsymbol w)$, $\mathcal{H}_1(\boldsymbol x), \mathcal{H}_2(\boldsymbol x)$ and $\mathcal{H}_3(\boldsymbol w, \boldsymbol x)$ as follows:
\begin{equation}\label{eq:h0_qubo}
	\mathcal{H}_0(\boldsymbol w) = \sum_{i=1}^{W}w_i^2,
\end{equation}
\begin{multline}\label{eq:h1_qubo}
	\mathcal{H}_1(\boldsymbol x) = \sum_{v=1}^{N_V}\left(1-\sum_{i=1}^{W}x_{v i}\right)^{2} = \\ 
	\sum_{v=1}^{N_V}\left(\left(\sum_{i=1}^{W}x_{v i}\right)^2 - 2\sum_{i=1}^{W}x_{v i}\right) + N_V = \\
	\sum_{v=1}^{N_V}\left(\sum_{i,j=1}^{W}x_{v i}x_{v j} - 2\sum_{i=1}^{W}x_{v i}^2\right) + N_V,
\end{multline}
\begin{multline}\label{eq:h2_qubo}
	\mathcal{H}_2(\boldsymbol x) = \sum_{(u,v)\in E}^{}\sum_{i=1}^{W} x_{ui}x_{vi} =  
	\sum_{u,v=1}^{N_V}\sum_{i=1}^{W}a_{u v}x_{u i}x_{v i},
\end{multline}
\begin{multline}\label{eq:h3_qubo}
	\mathcal{H}_3(\boldsymbol w, \boldsymbol x) = \sum_{(u,v)\in E}^{}\sum_{i=1}^{W}(1-w_i)\left(x_{ui}+x_{vi}\right) = \\
	\sum_{i=1}^{W}(1-w_i)\sum_{v=1}^{N_V}d_{v}x_{v i} = \\
	 -\sum_{v=1}^{N_V}\sum_{i=1}^{W}d_{v}w_{i}x_{v i} + \sum_{v=1}^{N_V}d_{v}\sum_{i=1}^{W}x_{v i}.
\end{multline}
In expanding the expression for $\mathcal{H}_1(\boldsymbol x)$ we exploit the fact that since $x_{vi}$ is binary then $x_{v i}^2 = x_{v i}$. 
Also we note that if $\mathcal{H}_1(\boldsymbol x)=0$ then the last term in $\mathcal{H}_3(\boldsymbol w, \boldsymbol x)$  equals $2N_E$.

Considering the equalities (\ref{eq:h0_qubo}-\ref{eq:h3_qubo}) for Hamiltonian terms $\mathcal{H}_0(\boldsymbol x), \mathcal{H}_1(\boldsymbol x), \mathcal{H}_2(\boldsymbol x)$ and $\mathcal{H}_3(\boldsymbol w, \boldsymbol x)$, 
we construct QUBO operator as block matrix as follows:
\begin{equation}
Q =
\begin{pmatrix}
Q_{1 1} & Q_{1 2} \\
Q_{2 1} & Q_{2 2}
\end{pmatrix},
\end{equation}
where
\begin{align}
    & Q_{1 1}=c_0 E_W, \\
    & Q_{1 2}=-\frac{c_2}{2}D \otimes E_W, \  Q_{2 1} = Q_{1 2}^T, \\
    & Q_{2 2}=c_1 E_{N_V} \otimes (I_W - 2E_W) + {c_1}A \otimes E_W.
\end{align}

Here $E_W$ denotes the identity matrix of size $W$, $I_W$ denotes a matrix with all elements equal to $1$ of those of size $W$, and $D=(d_1,\ldots,d_{N_V})$ is a row vector of graph vertex degrees.
We also employ the fact that terms of the form
\begin{equation}
	\sum_{u,v=1}^{N_V}\sum_{i,j=1}^{W} c_{u v}h_{i j}x_{u i}x_{v j}, 
\end{equation}
for some coefficients $c_{u v}=c_{v u}$ and $h_{i j} = h_{j i}$ can be represented by a quadratic form defined by Kronecker product $C \otimes H$, where $C$ and $H$ are matrices of  $c_{u v}$ and $h_{i j}$ correspondingly.
One can that matrix $Q$ is constructed so that $Q_{1 1}$ submatrix corresponds to the term $\mathcal{H}_0(\boldsymbol x)$ of Hamiltonian (\ref{hamiltonian:full}), $Q_{1 2}$ submatrix is for $\mathcal{H}_3(\boldsymbol w, \boldsymbol x)$ and $Q_{2 2}$ is for $\mathcal{H}_1(\boldsymbol x) + \mathcal{H}_2(\boldsymbol x)$.

It is worth to emphasize that it is the structure of encoding problem parameters into the spin vector, given by~\eqref{eq:notation_qubo}, that allow us to represent submatrices $Q_{1 2}$, $Q_{2 1}$, and $Q_{2 2}$ in the form of Kronecker products.
This feature of QUBO submatrices significantly speeds up their assembly using standard mathematical packages, e.g., ${\sf numpy}$ and ${\sf scipy}$.

\subsection{Dataset generation}{}\label{app:dataset}
We generate dataset are used binomial graphs~\cite{Batagelj2005Effic-5799}, or Erd{\"o}s-R\'enyi graphs, 
which have two parameters for generation: the number of nodes $N_V$ and the probability of an edge occurrence $p$. 
Each of possible $N=N_V\cdot (N_V-1)/2$ edges is chosen with probability $p$. 
Number of edges $N_E$ are drawn randomly from binomial distribution:
\begin{equation}
	P(N_E=x)=\binom{N}{x}p^x\cdot q^{(N-x)}.
\end{equation}

To take into account sparse and dense graphs, various probability $p$ options from 0.1 to 0.9 with an interval of 0.1 have been chosen, the number of graph nodes has been varied from 10 to 100 with a step of 10.
For each pair $(n, p)$, 10 connected graphs have been generated with different seed parameters. 
We note that disconnected graphs are not included the dataset. 
The overall characteristics of the dataset are given in the Table \ref{tab:dataset}.

\begin{table}[htp]
\vskip 3mm
\centering
\resizebox{\columnwidth}{!}{%
\begin{tabular}{|c|cc|cc|}
\hline
\multirow{2}{*}{\textbf{\begin{tabular}[c]{@{}c@{}}Number\\ of nodes\end{tabular}}} &
  \multicolumn{2}{c|}{\textbf{Number of edges}} &
  \multicolumn{2}{c|}{\textbf{QUBO matrix size}} \\ \cline{2-5} 
 &
  \multicolumn{1}{c|}{\textbf{$\quad$min$\quad$}} &
  \textbf{max} &
  \multicolumn{1}{c|}{\textbf{$\quad$min$\quad$}} &
  \textbf{max} \\ \hline
\textbf{10}  & \multicolumn{1}{c|}{9}   & 43   & \multicolumn{1}{c|}{44}  & 110  \\ \hline
\textbf{20}  & \multicolumn{1}{c|}{23}  & 176  & \multicolumn{1}{c|}{84}  & 315  \\ \hline
\textbf{30}  & \multicolumn{1}{c|}{39}  & 399  & \multicolumn{1}{c|}{124} & 589  \\ \hline
\textbf{40}  & \multicolumn{1}{c|}{74}  & 714  & \multicolumn{1}{c|}{205} & 943  \\ \hline
\textbf{50}  & \multicolumn{1}{c|}{118} & 1123 & \multicolumn{1}{c|}{255} & 1377 \\ \hline
\textbf{60}  & \multicolumn{1}{c|}{168} & 1625 & \multicolumn{1}{c|}{366} & 1891 \\ \hline
\textbf{70}  & \multicolumn{1}{c|}{231} & 2209 & \multicolumn{1}{c|}{426} & 2556 \\ \hline
\textbf{80}  & \multicolumn{1}{c|}{301} & 2879 & \multicolumn{1}{c|}{486} & 3321 \\ \hline
\textbf{90}  & \multicolumn{1}{c|}{372} & 3652 & \multicolumn{1}{c|}{546} & 4004 \\ \hline
\textbf{100} & \multicolumn{1}{c|}{470} & 4501 & \multicolumn{1}{c|}{707} & 4848 \\ \hline
\end{tabular}%
}
\caption{Characteristics of graph coloring dataset, the total number of instances is 900.}
\label{tab:dataset}
\end{table}

\subsection{Setting penalty values}\label{app:constraints}

Optimal penalty values guarantee the fulfilment of constraints for an optimal solution, but large values of $c_1$ and $c_2$ reduce the contribution of the initial objective function to the total energy and significantly increase the time to find the optimal solution. Our approach to solve this problem is as follows:
\begin{enumerate}
    \item Set the minimum possible penalty values $c_1$ and $c_2$ using trial runs, so that the contribution of the objective function is sufficient;
    \item Use all SimCIM iterations to select feasible solutions;  
    \item Take the feasible solution with the lowest energy.
\end{enumerate}

The following penalty values were set for the tests: 
\begin{equation}
    c_{0} = 1,\; c_{1} = 10 + p N_{V},\; c_{2} = 2.5.
\end{equation}

\subsection{Quantum-inspired annealing using SimCIM}

SimCIM~\cite{Tiunov2019} is an example of a quantum-inspired annealing algorithm, which works in an iterative manner.
SimCIM can be used for sampling low-energy spin configurations in the classical Ising model which Hamiltonian can be written as:
\begin{equation}
    \mathcal{H} = \sum_{i}h_is_i+\sum_{<i,j>}J_{ij}s_is_j,
\end{equation}
where $J$ represents the spin-spin interaction, $h$ represents the external field, and the $s_i$ are the individual spins on each of the lattice sites.
The Ising Hamiltonian can be directly transformed to a QUBO problem~\cite{Fedorov2021} and then quantum annealing can be applied to any optimization problem, which can be expressed into the QUBO form.
The SimCIM algorithm treats each spin value as a continuous variable $s_i\in[-1,1]$.
Each iteration of the algorithm starts with calculating the mean field of the following form:
\begin{equation}
	\Phi_i = \sum_{j \neq i}J_{ij}s_j + h_i,
\end{equation}
which act on each spin by all other spins.
Then the gradients for the spin values are calculated as follows:
\begin{equation}
	\Delta s_i = p_t s_i + \zeta \Phi_i + N(0,\sigma),
\end{equation}
where $p_t$ is a dynamic parameter dependent on SimCIM annealing process, overall feedforward factor $\zeta$ and  $N(0,\sigma)$ is a random variable sampled from the Gaussian distribution with zero mean and standard deviation $\sigma$.

Then the spin values are updated according to $s_i \leftarrow \phi(s_i + \Delta s_i)$, where $\phi(x)$ is the activation function
\begin{equation}\label{acti}
	\phi(x)=\begin{cases}
	x \textrm{ for } |x|\leq 1;\\
	x/|x|, \textrm{ otherwise.}
\end{cases}
\end{equation}

After multiple updates, the spins will tend to either $-1$ or $+1$ and the final discrete spin configuration is obtained by taking the sign of each $s_i$.

In our implementation we add several improvements to SimCIM algorithm defined in the original paper~\cite{Tiunov2019}. 
In particular, we normalize the value of the Gaussian noise to gradient norm and introduced gradient quantization, which made the solver more stable near optimum points.

\section*{Appendix}\label{sec:appendix}

\begin{table}[H]
\centering
\resizebox{\columnwidth}{!}{%
\begin{tabular}{|c|cccc}
\hline
\multirow{2}{*}{\textbf{\begin{tabular}[c]{@{}c@{}}Number\\ of nodes\end{tabular}}} & 
\multicolumn{2}{c|}{\textbf{\begin{tabular}[c]{@{}c@{}}Original QUBO\\ transformation\end{tabular}}} 
& \multicolumn{2}{c|}{\textbf{\begin{tabular}[c]{@{}c@{}}Proposed QUBO\\ transformation\end{tabular}}} \\ \cline{2-5} 
 & \multicolumn{1}{l|}{Number of colors} & \multicolumn{1}{l|}{Run time} & \multicolumn{1}{l|}{Number of colors} & \multicolumn{1}{l|}{Run time} \\ \hline
\textbf{10} & \multicolumn{1}{c|}{4.34} & \multicolumn{1}{c|}{0.28} & \multicolumn{1}{c|}{4.34} & \multicolumn{1}{c|}{0.19} \\ \hline
\textbf{20} & \multicolumn{1}{c|}{6.47} & \multicolumn{1}{c|}{0.62} & \multicolumn{1}{c|}{6.36} & \multicolumn{1}{c|}{0.45} \\ \hline
\textbf{30} & \multicolumn{1}{c|}{8.24} & \multicolumn{1}{c|}{7.67} & \multicolumn{1}{c|}{8.02} & \multicolumn{1}{c|}{4.95} \\ \hline
\textbf{40} & \multicolumn{1}{c|}{10.31} & \multicolumn{1}{c|}{14.22} & \multicolumn{1}{c|}{9.39} & \multicolumn{1}{c|}{8.90} \\ \hline
\textbf{50} & \multicolumn{1}{c|}{12.41} & \multicolumn{1}{c|}{26.28} & \multicolumn{1}{c|}{10.96} & \multicolumn{1}{c|}{16.82} \\ \hline
\textbf{60} & \multicolumn{1}{c|}{14.53} & \multicolumn{1}{c|}{42.01} & \multicolumn{1}{c|}{12.44} & \multicolumn{1}{c|}{28.51} \\ \hline
\textbf{70} & \multicolumn{1}{c|}{16.52} & \multicolumn{1}{c|}{63.89} & \multicolumn{1}{c|}{14.01} & \multicolumn{1}{c|}{61.58} \\ \hline
\textbf{80} & \multicolumn{1}{c|}{18.03} & \multicolumn{1}{c|}{98.50} & \multicolumn{1}{c|}{15.56} & \multicolumn{1}{c|}{69.00} \\ \hline
\textbf{90} & \multicolumn{1}{c|}{19.74} & \multicolumn{1}{c|}{106.61} & \multicolumn{1}{c|}{17.02} & \multicolumn{1}{c|}{79.87} \\ \hline
\textbf{100} & \multicolumn{1}{c|}{20.65} & \multicolumn{1}{c|}{140.41} & \multicolumn{1}{c|}{18.54} & \multicolumn{1}{c|}{123.13} \\ \hline
\end{tabular}%
}
\\ \textbf{Average results \\ (lower is better)}
\caption{Comparison of proposed QUBO transformation for graph coloring problem to Original QUBO transformation described in Ref.~\cite{Lucas2014}.Experiments were performed on the same dataset of 900 randomly generated graphs with the use of SimCIM. Results shows that the proposed QUBO runs faster, giving on average lower or the same number of colors.
\label{tab:qubocomparision}}
\end{table}

\begin{widetext}
\section*{Supplementary Material}

\subsection*{Number of colors}

\begin{longtable}{|l|r|r|rr|cc|rr|rr|}
\caption{Numerical results obtained with largest degree first (LDF) heuristics, open-source mixed integer programming solver (GLPK), Gurobi optimization software, and SimCIM quantum-inspired optimization on number of colors averaged by 10 graph with different number of nodes ($N_V$) and edge probability ($p$). The best result is highlighted in bold. SimCIM shows the best results in the $p$ range $[0.3,\cdots, 0.7]$.}
\label{tab:benchmark_nc}\\
\hline
\multirow{2}{*}{} & \multicolumn{1}{c|}{\multirow{2}{*}{\textbf{$N_V$}}} & \multicolumn{1}{c|}{\multirow{2}{*}{\textbf{$p$}}} & \multicolumn{2}{c|}{\textbf{LDF}} & \multicolumn{2}{c|}{\textbf{GLPK}} & \multicolumn{2}{c|}{\textbf{Gurobi}} & \multicolumn{2}{c|}{\textbf{SimCIM}} \\ \cline{4-11} 
 & \multicolumn{1}{c|}{} & \multicolumn{1}{c|}{} & \multicolumn{1}{c|}{\textbf{mean}} & \multicolumn{1}{c|}{\textbf{std}} & \multicolumn{1}{c|}{\textbf{mean}} & \textbf{std} & \multicolumn{1}{c|}{\textbf{mean}} & \multicolumn{1}{c|}{\textbf{std}} & \multicolumn{1}{c|}{\textbf{mean}} & \multicolumn{1}{c|}{\textbf{std}} \\ \hline
\textbf{1} & 10 & 0.1 & \multicolumn{1}{r|}{3.0} & 0.000 & \multicolumn{1}{r|}{2.50} & \multicolumn{1}{r|}{0.527} & \multicolumn{1}{r|}{2.5} & 0.527 & \multicolumn{1}{r|}{2.5} & 0.527 \\ \hline
\textbf{2} & 10 & 0.2 & \multicolumn{1}{r|}{3.0} & 0.000 & \multicolumn{1}{r|}{3.00} & \multicolumn{1}{r|}{0.000} & \multicolumn{1}{r|}{3.0} & 0.000 & \multicolumn{1}{r|}{3.0} & 0.000 \\ \hline
\textbf{3} & 10 & 0.3 & \multicolumn{1}{r|}{3.1} & 0.316 & \multicolumn{1}{r|}{3.10} & \multicolumn{1}{r|}{0.316} & \multicolumn{1}{r|}{3.1} & 0.316 & \multicolumn{1}{r|}{3.1} & 0.316 \\ \hline
\textbf{4} & 10 & 0.4 & \multicolumn{1}{r|}{3.6} & 0.516 & \multicolumn{1}{r|}{3.50} & \multicolumn{1}{r|}{0.527} & \multicolumn{1}{r|}{3.5} & 0.527 & \multicolumn{1}{r|}{3.5} & 0.527 \\ \hline
\textbf{5} & 10 & 0.5 & \multicolumn{1}{r|}{4.0} & 0.471 & \multicolumn{1}{r|}{4.00} & \multicolumn{1}{r|}{0.471} & \multicolumn{1}{r|}{4.0} & 0.471 & \multicolumn{1}{r|}{4.0} & 0.471 \\ \hline
\textbf{6} & 10 & 0.6 & \multicolumn{1}{r|}{4.7} & 0.675 & \multicolumn{1}{r|}{4.50} & \multicolumn{1}{r|}{0.527} & \multicolumn{1}{r|}{4.5} & 0.527 & \multicolumn{1}{r|}{4.5} & 0.527 \\ \hline
\textbf{7} & 10 & 0.7 & \multicolumn{1}{r|}{5.2} & 0.422 & \multicolumn{1}{r|}{5.00} & \multicolumn{1}{r|}{0.667} & \multicolumn{1}{r|}{5.0} & 0.667 & \multicolumn{1}{r|}{5.0} & 0.667 \\ \hline
\textbf{8} & 10 & 0.8 & \multicolumn{1}{r|}{5.9} & 0.994 & \multicolumn{1}{r|}{5.90} & \multicolumn{1}{r|}{0.994} & \multicolumn{1}{r|}{5.9} & 0.994 & \multicolumn{1}{r|}{5.9} & 0.994 \\ \hline
\textbf{9} & 10 & 0.9 & \multicolumn{1}{r|}{7.6} & 0.843 & \multicolumn{1}{r|}{7.60} & \multicolumn{1}{r|}{0.843} & \multicolumn{1}{r|}{7.6} & 0.843 & \multicolumn{1}{r|}{7.6} & 0.843 \\ \hline
\textbf{10} & 20 & 0.1 & \multicolumn{1}{r|}{3.0} & 0.000 & \multicolumn{1}{r|}{3.00} & \multicolumn{1}{r|}{0.000} & \multicolumn{1}{r|}{3.0} & 0.000 & \multicolumn{1}{r|}{3.0} & 0.000 \\ \hline
\textbf{11} & 20 & 0.2 & \multicolumn{1}{r|}{3.7} & 0.675 & \multicolumn{1}{r|}{3.50} & \multicolumn{1}{r|}{0.527} & \multicolumn{1}{r|}{3.5} & 0.527 & \multicolumn{1}{r|}{3.5} & 0.527 \\ \hline
\textbf{12} & 20 & 0.3 & \multicolumn{1}{r|}{4.4} & 0.516 & \multicolumn{1}{r|}{4.10} & \multicolumn{1}{r|}{0.316} & \multicolumn{1}{r|}{4.1} & 0.316 & \multicolumn{1}{r|}{4.1} & 0.316 \\ \hline
\textbf{13} & 20 & 0.4 & \multicolumn{1}{r|}{5.6} & 0.516 & \multicolumn{1}{r|}{5.00} & \multicolumn{1}{r|}{0.000} & \multicolumn{1}{r|}{5.0} & 0.000 & \multicolumn{1}{r|}{5.0} & 0.000 \\ \hline
\textbf{14} & 20 & 0.5 & \multicolumn{1}{r|}{6.3} & 0.949 & \multicolumn{1}{r|}{5.70} & \multicolumn{1}{r|}{0.483} & \multicolumn{1}{r|}{5.7} & 0.483 & \multicolumn{1}{r|}{5.7} & 0.483 \\ \hline
\textbf{15} & 20 & 0.6 & \multicolumn{1}{r|}{7.5} & 0.527 & \multicolumn{1}{r|}{6.80} & \multicolumn{1}{r|}{0.422} & \multicolumn{1}{r|}{6.8} & 0.422 & \multicolumn{1}{r|}{6.8} & 0.422 \\ \hline
\textbf{16} & 20 & 0.7 & \multicolumn{1}{r|}{8.4} & 0.843 & \multicolumn{1}{r|}{7.40} & \multicolumn{1}{r|}{0.516} & \multicolumn{1}{r|}{7.4} & 0.516 & \multicolumn{1}{r|}{7.4} & 0.516 \\ \hline
\textbf{17} & 20 & 0.8 & \multicolumn{1}{r|}{10.0} & 0.816 & \multicolumn{1}{r|}{9.20} & \multicolumn{1}{r|}{0.632} & \multicolumn{1}{r|}{9.2} & 0.632 & \multicolumn{1}{r|}{9.2} & 0.632 \\ \hline
\textbf{18} & 20 & 0.9 & \multicolumn{1}{r|}{12.5} & 0.850 & \multicolumn{1}{r|}{12.50} & \multicolumn{1}{r|}{0.850} & \multicolumn{1}{r|}{12.5} & 0.850 & \multicolumn{1}{r|}{12.5} & 0.850 \\ \hline
\textbf{19} & 30 & 0.1 & \multicolumn{1}{r|}{3.6} & 0.516 & \multicolumn{1}{r|}{3.00} & \multicolumn{1}{r|}{0.000} & \multicolumn{1}{r|}{3.0} & 0.000 & \multicolumn{1}{r|}{3.0} & 0.000 \\ \hline
\textbf{20} & 30 & 0.2 & \multicolumn{1}{r|}{4.7} & 0.483 & \multicolumn{1}{r|}{4.00} & \multicolumn{1}{r|}{0.000} & \multicolumn{1}{r|}{4.0} & 0.000 & \multicolumn{1}{r|}{4.0} & 0.000 \\ \hline
\textbf{21} & 30 & 0.3 & \multicolumn{1}{r|}{6.0} & 0.000 & \multicolumn{1}{r|}{5.00} & \multicolumn{1}{r|}{0.000} & \multicolumn{1}{r|}{5.0} & 0.000 & \multicolumn{1}{r|}{5.0} & 0.000 \\ \hline
\textbf{22} & 30 & 0.4 & \multicolumn{1}{r|}{7.2} & 0.422 & \multicolumn{1}{r|}{6.00} & \multicolumn{1}{r|}{0.000} & \multicolumn{1}{r|}{6.0} & 0.000 & \multicolumn{1}{r|}{6.0} & 0.000 \\ \hline
\textbf{23} & 30 & 0.5 & \multicolumn{1}{r|}{8.3} & 0.675 & \multicolumn{1}{r|}{7.20} & \multicolumn{1}{r|}{0.422} & \multicolumn{1}{r|}{7.2} & 0.422 & \multicolumn{1}{r|}{7.2} & 0.422 \\ \hline
\textbf{24} & 30 & 0.6 & \multicolumn{1}{r|}{9.8} & 0.919 & \multicolumn{1}{r|}{8.60} & \multicolumn{1}{r|}{0.516} & \multicolumn{1}{r|}{8.5} & 0.527 & \multicolumn{1}{r|}{8.5} & 0.527 \\ \hline
\textbf{25} & 30 & 0.7 & \multicolumn{1}{r|}{11.5} & 0.850 & \multicolumn{1}{r|}{10.50} & \multicolumn{1}{r|}{0.972} & \multicolumn{1}{r|}{10.1} & 0.738 & \multicolumn{1}{r|}{10.1} & 0.738 \\ \hline
\textbf{26} & 30 & 0.8 & \multicolumn{1}{r|}{13.7} & 0.949 & \multicolumn{1}{r|}{13.40} & \multicolumn{1}{r|}{0.966} & \multicolumn{1}{r|}{12.6} & 0.699 & \multicolumn{1}{r|}{12.6} & 0.699 \\ \hline
\textbf{27} & 30 & 0.9 & \multicolumn{1}{r|}{16.5} & 1.509 & \multicolumn{1}{r|}{16.25} & \multicolumn{1}{r|}{1.669} & \multicolumn{1}{r|}{15.8} & 1.398 & \multicolumn{1}{r|}{15.8} & 1.398 \\ \hline
\textbf{28} & 40 & 0.1 & \multicolumn{1}{r|}{4.0} & 0.000 & \multicolumn{1}{c|}{-} & - & \multicolumn{1}{r|}{3.1} & 0.316 & \multicolumn{1}{r|}{3.1} & 0.316 \\ \hline
\textbf{29} & 40 & 0.2 & \multicolumn{1}{r|}{5.3} & 0.483 & \multicolumn{1}{c|}{-} & - & \multicolumn{1}{r|}{4.5} & 0.527 & \multicolumn{1}{r|}{4.5} & 0.527 \\ \hline
\textbf{30} & 40 & 0.3 & \multicolumn{1}{r|}{7.1} & 0.568 & \multicolumn{1}{c|}{-} & - & \multicolumn{1}{r|}{6.0} & 0.000 & \multicolumn{1}{r|}{6.0} & 0.000 \\ \hline
\textbf{31} & 40 & 0.4 & \multicolumn{1}{r|}{8.5} & 0.527 & \multicolumn{1}{c|}{-} & - & \multicolumn{1}{r|}{6.9} & 0.316 & \multicolumn{1}{r|}{6.9} & 0.316 \\ \hline
\textbf{32} & 40 & 0.5 & \multicolumn{1}{r|}{10.1} & 0.738 & \multicolumn{1}{c|}{-} & - & \multicolumn{1}{r|}{\textit{\textbf{8.3}}} & 0.483 & \multicolumn{1}{r|}{8.4} & 0.516 \\ \hline
\textbf{33} & 40 & 0.6 & \multicolumn{1}{r|}{12.3} & 0.823 & \multicolumn{1}{c|}{-} & - & \multicolumn{1}{r|}{9.9} & 0.316 & \multicolumn{1}{r|}{9.9} & 0.316 \\ \hline
\textbf{34} & 40 & 0.7 & \multicolumn{1}{r|}{13.8} & 0.919 & \multicolumn{1}{c|}{-} & - & \multicolumn{1}{r|}{11.8} & 0.632 & \multicolumn{1}{r|}{11.8} & 0.632 \\ \hline
\textbf{35} & 40 & 0.8 & \multicolumn{1}{r|}{16.9} & 0.738 & \multicolumn{1}{c|}{-} & - & \multicolumn{1}{r|}{\textit{\textbf{14.9}}} & 0.738 & \multicolumn{1}{r|}{15.0} & 0.816 \\ \hline
\textbf{36} & 40 & 0.9 & \multicolumn{1}{r|}{20.3} & 1.160 & \multicolumn{1}{c|}{-} & - & \multicolumn{1}{r|}{19.0} & 0.943 & \multicolumn{1}{r|}{19.0} & 0.943 \\ \hline
\textbf{37} & 50 & 0.1 & \multicolumn{1}{r|}{4.4} & 0.516 & \multicolumn{1}{c|}{-} & - & \multicolumn{1}{r|}{4.0} & 0.000 & \multicolumn{1}{r|}{4.0} & 0.000 \\ \hline
\textbf{38} & 50 & 0.2 & \multicolumn{1}{r|}{6.2} & 0.422 & \multicolumn{1}{c|}{-} & - & \multicolumn{1}{r|}{5.0} & 0.000 & \multicolumn{1}{r|}{5.0} & 0.000 \\ \hline
\textbf{39} & 50 & 0.3 & \multicolumn{1}{r|}{7.8} & 0.422 & \multicolumn{1}{c|}{-} & - & \multicolumn{1}{r|}{\textit{\textbf{6.5}}} & 0.527 & \multicolumn{1}{r|}{6.7} & 0.483 \\ \hline
\textbf{40} & 50 & 0.4 & \multicolumn{1}{r|}{10.2} & 0.422 & \multicolumn{1}{c|}{-} & - & \multicolumn{1}{r|}{7.9} & 0.316 & \multicolumn{1}{r|}{7.9} & 0.316 \\ \hline
\textbf{41} & 50 & 0.5 & \multicolumn{1}{r|}{11.9} & 0.568 & \multicolumn{1}{c|}{-} & - & \multicolumn{1}{r|}{9.8} & 0.422 & \multicolumn{1}{r|}{\textit{\textbf{9.7}}} & 0.483 \\ \hline
\textbf{42} & 50 & 0.6 & \multicolumn{1}{r|}{14.1} & 0.568 & \multicolumn{1}{c|}{-} & - & \multicolumn{1}{r|}{\textit{\textbf{11.3}}} & 0.483 & \multicolumn{1}{r|}{11.5} & 0.527 \\ \hline
\textbf{43} & 50 & 0.7 & \multicolumn{1}{r|}{16.2} & 0.789 & \multicolumn{1}{c|}{-} & - & \multicolumn{1}{r|}{\textit{\textbf{13.8}}} & 0.632 & \multicolumn{1}{r|}{14.2} & 0.632 \\ \hline
\textbf{44} & 50 & 0.8 & \multicolumn{1}{r|}{20.2} & 0.632 & \multicolumn{1}{c|}{-} & - & \multicolumn{1}{r|}{\textit{\textbf{17.1}}} & 0.568 & \multicolumn{1}{r|}{17.2} & 0.422 \\ \hline
\textbf{45} & 50 & 0.9 & \multicolumn{1}{r|}{24.2} & 1.229 & \multicolumn{1}{c|}{-} & - & \multicolumn{1}{r|}{22.5} & 1.179 & \multicolumn{1}{r|}{22.5} & 1.179 \\ \hline
\textbf{46} & 60 & 0.1 & \multicolumn{1}{r|}{5.3} & 0.483 & \multicolumn{1}{c|}{-} & - & \multicolumn{1}{r|}{4.0} & 0.000 & \multicolumn{1}{r|}{4.0} & 0.000 \\ \hline
\textbf{47} & 60 & 0.2 & \multicolumn{1}{r|}{7.3} & 0.483 & \multicolumn{1}{c|}{-} & - & \multicolumn{1}{r|}{\textit{\textbf{5.7}}} & 0.483 & \multicolumn{1}{r|}{5.8} & 0.422 \\ \hline
\textbf{48} & 60 & 0.3 & \multicolumn{1}{r|}{8.7} & 0.675 & \multicolumn{1}{c|}{-} & - & \multicolumn{1}{r|}{7.0} & 0.000 & \multicolumn{1}{r|}{7.0} & 0.000 \\ \hline
\textbf{49} & 60 & 0.4 & \multicolumn{1}{r|}{11.4} & 0.699 & \multicolumn{1}{c|}{-} & - & \multicolumn{1}{r|}{9.0} & 0.000 & \multicolumn{1}{r|}{9.0} & 0.000 \\ \hline
\textbf{50} & 60 & 0.5 & \multicolumn{1}{r|}{13.8} & 0.632 & \multicolumn{1}{c|}{-} & - & \multicolumn{1}{r|}{10.9} & 0.316 & \multicolumn{1}{r|}{10.9} & 0.316 \\ \hline
\textbf{51} & 60 & 0.6 & \multicolumn{1}{r|}{16.3} & 1.160 & \multicolumn{1}{c|}{-} & - & \multicolumn{1}{r|}{12.9} & 0.316 & \multicolumn{1}{r|}{12.9} & 0.316 \\ \hline
\textbf{52} & 60 & 0.7 & \multicolumn{1}{r|}{18.8} & 1.135 & \multicolumn{1}{c|}{-} & - & \multicolumn{1}{r|}{\textit{\textbf{15.8}}} & 0.422 & \multicolumn{1}{r|}{16.1} & 0.568 \\ \hline
\textbf{53} & 60 & 0.8 & \multicolumn{1}{r|}{23.2} & 0.789 & \multicolumn{1}{c|}{-} & - & \multicolumn{1}{r|}{\textit{\textbf{19.4}}} & 0.516 & \multicolumn{1}{r|}{19.9} & 0.568 \\ \hline
\textbf{54} & 60 & 0.9 & \multicolumn{1}{r|}{28.7} & 1.059 & \multicolumn{1}{c|}{-} & - & \multicolumn{1}{r|}{\textit{\textbf{25.8}}} & 0.789 & \multicolumn{1}{r|}{26.0} & 0.816 \\ \hline
\textbf{55} & 70 & 0.1 & \multicolumn{1}{r|}{5.4} & 0.516 & \multicolumn{1}{c|}{-} & - & \multicolumn{1}{r|}{4.0} & 0.000 & \multicolumn{1}{r|}{4.0} & 0.000 \\ \hline
\textbf{56} & 70 & 0.2 & \multicolumn{1}{r|}{7.7} & 0.675 & \multicolumn{1}{c|}{-} & - & \multicolumn{1}{r|}{6.0} & 0.000 & \multicolumn{1}{r|}{6.0} & 0.000 \\ \hline
\textbf{57} & 70 & 0.3 & \multicolumn{1}{r|}{9.7} & 0.675 & \multicolumn{1}{c|}{-} & - & \multicolumn{1}{r|}{8.0} & 0.000 & \multicolumn{1}{r|}{\textit{\textbf{7.9}}} & 0.316 \\ \hline
\textbf{58} & 70 & 0.4 & \multicolumn{1}{r|}{12.3} & 0.675 & \multicolumn{1}{c|}{-} & - & \multicolumn{1}{r|}{\textit{\textbf{10.1}}} & 0.316 & \multicolumn{1}{r|}{10.5} & 1.269 \\ \hline
\textbf{59} & 70 & 0.5 & \multicolumn{1}{r|}{15.2} & 0.632 & \multicolumn{1}{c|}{-} & - & \multicolumn{1}{r|}{\textit{\textbf{12.3}}} & 0.483 & \multicolumn{1}{r|}{12.5} & 0.527 \\ \hline
\textbf{60} & 70 & 0.6 & \multicolumn{1}{r|}{18.4} & 0.516 & \multicolumn{1}{c|}{-} & - & \multicolumn{1}{r|}{14.9} & 0.316 & \multicolumn{1}{r|}{\textit{\textbf{14.6}}} & 0.699 \\ \hline
\textbf{61} & 70 & 0.7 & \multicolumn{1}{r|}{22.0} & 1.247 & \multicolumn{1}{c|}{-} & - & \multicolumn{1}{r|}{\textit{\textbf{17.8}}} & 0.422 & \multicolumn{1}{r|}{17.9} & 0.316 \\ \hline
\textbf{62} & 70 & 0.8 & \multicolumn{1}{r|}{26.4} & 1.713 & \multicolumn{1}{c|}{-} & - & \multicolumn{1}{r|}{\textit{\textbf{21.6}}} & 0.699 & \multicolumn{1}{r|}{22.1} & 0.568 \\ \hline
\textbf{63} & 70 & 0.9 & \multicolumn{1}{r|}{32.5} & 1.269 & \multicolumn{1}{c|}{-} & - & \multicolumn{1}{r|}{\textit{\textbf{28.6}}} & 0.966 & \multicolumn{1}{r|}{29.2} & 0.789 \\ \hline
\textbf{64} & 80 & 0.1 & \multicolumn{1}{r|}{5.7} & 0.483 & \multicolumn{1}{c|}{-} & - & \multicolumn{1}{r|}{\textit{\textbf{4.3}}} & 0.483 & \multicolumn{1}{r|}{4.8} & 0.422 \\ \hline
\textbf{65} & 80 & 0.2 & \multicolumn{1}{r|}{8.3} & 0.483 & \multicolumn{1}{c|}{-} & - & \multicolumn{1}{r|}{\textit{\textbf{6.4}}} & 0.516 & \multicolumn{1}{r|}{6.8} & 0.422 \\ \hline
\textbf{66} & 80 & 0.3 & \multicolumn{1}{r|}{11.2} & 0.632 & \multicolumn{1}{c|}{-} & - & \multicolumn{1}{r|}{8.9} & 0.316 & \multicolumn{1}{r|}{8.9} & 0.316 \\ \hline
\textbf{67} & 80 & 0.4 & \multicolumn{1}{r|}{13.5} & 0.527 & \multicolumn{1}{c|}{-} & - & \multicolumn{1}{r|}{11.6} & 0.516 & \multicolumn{1}{r|}{\textit{\textbf{11.0}}} & 0.000 \\ \hline
\textbf{68} & 80 & 0.5 & \multicolumn{1}{r|}{16.9} & 0.994 & \multicolumn{1}{c|}{-} & - & \multicolumn{1}{r|}{14.2} & 0.632 & \multicolumn{1}{r|}{\textit{\textbf{13.6}}} & 0.516 \\ \hline
\textbf{69} & 80 & 0.6 & \multicolumn{1}{r|}{20.5} & 0.972 & \multicolumn{1}{c|}{-} & - & \multicolumn{1}{r|}{\textit{\textbf{16.6}}} & 0.516 & \multicolumn{1}{r|}{16.9} & 1.853 \\ \hline
\textbf{70} & 80 & 0.7 & \multicolumn{1}{r|}{23.5} & 0.972 & \multicolumn{1}{c|}{-} & - & \multicolumn{1}{r|}{\textit{\textbf{20.0}}} & 0.667 & \multicolumn{1}{r|}{20.2} & 0.632 \\ \hline
\textbf{71} & 80 & 0.8 & \multicolumn{1}{r|}{29.5} & 1.650 & \multicolumn{1}{c|}{-} & - & \multicolumn{1}{r|}{\textit{\textbf{24.6}}} & 0.699 & \multicolumn{1}{r|}{24.9} & 0.738 \\ \hline
\textbf{72} & 80 & 0.9 & \multicolumn{1}{r|}{36.6} & 2.119 & \multicolumn{1}{c|}{-} & - & \multicolumn{1}{r|}{\textit{\textbf{31.5}}} & 1.179 & \multicolumn{1}{r|}{32.5} & 1.354 \\ \hline
\textbf{73} & 90 & 0.1 & \multicolumn{1}{r|}{5.9} & 0.316 & \multicolumn{1}{c|}{-} & - & \multicolumn{1}{r|}{\textit{\textbf{4.9}}} & 0.316 & \multicolumn{1}{r|}{5.0} & 0.000 \\ \hline
\textbf{74} & 90 & 0.2 & \multicolumn{1}{r|}{9.2} & 0.632 & \multicolumn{1}{c|}{-} & - & \multicolumn{1}{r|}{7.0} & 0.000 & \multicolumn{1}{r|}{7.0} & 0.000 \\ \hline
\textbf{75} & 90 & 0.3 & \multicolumn{1}{r|}{12.2} & 0.632 & \multicolumn{1}{c|}{-} & - & \multicolumn{1}{r|}{9.7} & 0.483 & \multicolumn{1}{r|}{\textit{\textbf{9.5}}} & 0.527 \\ \hline
\textbf{76} & 90 & 0.4 & \multicolumn{1}{r|}{15.2} & 0.632 & \multicolumn{1}{c|}{-} & - & \multicolumn{1}{r|}{12.9} & 0.316 & \multicolumn{1}{r|}{\textit{\textbf{12.0}}} & 0.471 \\ \hline
\textbf{77} & 90 & 0.5 & \multicolumn{1}{r|}{18.3} & 0.949 & \multicolumn{1}{c|}{-} & - & \multicolumn{1}{r|}{15.6} & 0.699 & \multicolumn{1}{r|}{\textit{\textbf{15.1}}} & 0.568 \\ \hline
\textbf{78} & 90 & 0.6 & \multicolumn{1}{r|}{22.2} & 0.919 & \multicolumn{1}{c|}{-} & - & \multicolumn{1}{r|}{19.8} & 0.789 & \multicolumn{1}{r|}{\textit{\textbf{18.3}}} & 0.483 \\ \hline
\textbf{79} & 90 & 0.7 & \multicolumn{1}{r|}{26.0} & 1.333 & \multicolumn{1}{c|}{-} & - & \multicolumn{1}{r|}{22.9} & 1.197 & \multicolumn{1}{r|}{\textit{\textbf{22.6}}} & 0.516 \\ \hline
\textbf{80} & 90 & 0.8 & \multicolumn{1}{r|}{31.8} & 0.789 & \multicolumn{1}{c|}{-} & - & \multicolumn{1}{r|}{\textit{\textbf{27.5}}} & 1.080 & \multicolumn{1}{r|}{27.6} & 0.516 \\ \hline
\textbf{81} & 90 & 0.9 & \multicolumn{1}{r|}{40.1} & 2.132 & \multicolumn{1}{c|}{-} & - & \multicolumn{1}{r|}{\textit{\textbf{34.6}}} & 0.843 & \multicolumn{1}{r|}{36.3} & 1.160 \\ \hline
\textbf{82} & 100 & 0.1 & \multicolumn{1}{r|}{6.6} & 0.516 & \multicolumn{1}{c|}{-} & - & \multicolumn{1}{r|}{5.0} & 0.000 & \multicolumn{1}{r|}{5.0} & 0.000 \\ \hline
\textbf{83} & 100 & 0.2 & \multicolumn{1}{r|}{10.1} & 0.568 & \multicolumn{1}{c|}{-} & - & \multicolumn{1}{r|}{\textit{\textbf{7.5}}} & 0.527 & \multicolumn{1}{r|}{7.9} & 0.316 \\ \hline
\textbf{84} & 100 & 0.3 & \multicolumn{1}{r|}{13.0} & 0.471 & \multicolumn{1}{c|}{-} & - & \multicolumn{1}{r|}{11.1} & 0.316 & \multicolumn{1}{r|}{\textit{\textbf{10.2}}} & 0.422 \\ \hline
\textbf{85} & 100 & 0.4 & \multicolumn{1}{r|}{16.5} & 0.850 & \multicolumn{1}{c|}{-} & - & \multicolumn{1}{r|}{14.5} & 0.850 & \multicolumn{1}{r|}{\textit{\textbf{13.2}}} & 0.422 \\ \hline
\textbf{86} & 100 & 0.5 & \multicolumn{1}{r|}{19.9} & 0.738 & \multicolumn{1}{c|}{-} & - & \multicolumn{1}{r|}{18.4} & 0.699 & \multicolumn{1}{r|}{\textit{\textbf{16.2}}} & 0.422 \\ \hline
\textbf{87} & 100 & 0.6 & \multicolumn{1}{r|}{24.1} & 0.568 & \multicolumn{1}{c|}{-} & - & \multicolumn{1}{r|}{22.5} & 0.707 & \multicolumn{1}{r|}{\textit{\textbf{19.9}}} & 0.568 \\ \hline
\textbf{88} & 100 & 0.7 & \multicolumn{1}{r|}{28.5} & 1.269 & \multicolumn{1}{c|}{-} & - & \multicolumn{1}{r|}{27.0} & 1.054 & \multicolumn{1}{r|}{\textit{\textbf{24.2}}} & 0.632 \\ \hline
\textbf{89} & 100 & 0.8 & \multicolumn{1}{r|}{35.1} & 1.370 & \multicolumn{1}{c|}{-} & - & \multicolumn{1}{r|}{32.7} & 1.337 & \multicolumn{1}{r|}{\textit{\textbf{30.5}}} & 0.850 \\ \hline
\textbf{90} & 100 & 0.9 & \multicolumn{1}{r|}{44.3} & 1.703 & \multicolumn{1}{c|}{-} & - & \multicolumn{1}{r|}{\textit{\textbf{38.1}}} & 0.738 & \multicolumn{1}{r|}{40.2} & 0.919 \\ \hline
\end{longtable}
\centering\textbf{Average number of colors \\ (lower is better)}

\subsection*{Time to solution}
\begin{longtable}{|l|r|r|rr|cc|rr|rr|}
\caption{Mean time to solution (seconds) depends on number of nodes ($N_V$) and edge probability ($p$) for open source solver GLPK, Gurobi and quantum inspired SimCIM. The best result in average number of colors and time to solution is highlighted in bold.}\\
\hline
\multirow{2}{*}{} & \multicolumn{1}{c|}{\multirow{2}{*}{\textbf{$N_V$}}} & \multicolumn{1}{c|}{\multirow{2}{*}{\textbf{$p$}}} & \multicolumn{2}{c|}{\textbf{GLPK}} & \multicolumn{2}{c|}{\textbf{Gurobi}} & \multicolumn{2}{c|}{\textbf{SimCIM}} \\ \cline{4-9} 
 & \multicolumn{1}{c|}{} & \multicolumn{1}{c|}{} & \multicolumn{1}{c|}{\textbf{mean}} & \textbf{std} & \multicolumn{1}{c|}{\textbf{mean}} & \multicolumn{1}{c|}{\textbf{std}} & \multicolumn{1}{c|}{\textbf{mean}} & \multicolumn{1}{c|}{\textbf{std}} \\ \hline
\textbf{1} & 10 & 0.1 & \multicolumn{1}{r|}{0.001} & \multicolumn{1}{r|}{0.001} & \multicolumn{1}{r|}{0.001} & 0.000 & \multicolumn{1}{r|}{0.198} & 0.056 \\ \hline
\textbf{2} & 10 & 0.2 & \multicolumn{1}{r|}{0.002} & \multicolumn{1}{r|}{0.000} & \multicolumn{1}{r|}{0.001} & 0.000 & \multicolumn{1}{r|}{0.232} & 0.007 \\ \hline
\textbf{3} & 10 & 0.3 & \multicolumn{1}{r|}{0.003} & \multicolumn{1}{r|}{0.002} & \multicolumn{1}{r|}{0.001} & 0.000 & \multicolumn{1}{r|}{0.238} & 0.015 \\ \hline
\textbf{4} & 10 & 0.4 & \multicolumn{1}{r|}{0.007} & \multicolumn{1}{r|}{0.004} & \multicolumn{1}{r|}{0.001} & 0.000 & \multicolumn{1}{r|}{0.229} & 0.021 \\ \hline
\textbf{5} & 10 & 0.5 & \multicolumn{1}{r|}{0.012} & \multicolumn{1}{r|}{0.006} & \multicolumn{1}{r|}{0.001} & 0.000 & \multicolumn{1}{r|}{0.222} & 0.011 \\ \hline
\textbf{6} & 10 & 0.6 & \multicolumn{1}{r|}{0.032} & \multicolumn{1}{r|}{0.022} & \multicolumn{1}{r|}{0.002} & 0.000 & \multicolumn{1}{r|}{0.239} & 0.022 \\ \hline
\textbf{7} & 10 & 0.7 & \multicolumn{1}{r|}{0.080} & \multicolumn{1}{r|}{0.073} & \multicolumn{1}{r|}{0.002} & 0.001 & \multicolumn{1}{r|}{0.265} & 0.044 \\ \hline
\textbf{8} & 10 & 0.8 & \multicolumn{1}{r|}{1.907} & \multicolumn{1}{r|}{4.984} & \multicolumn{1}{r|}{0.002} & 0.001 & \multicolumn{1}{r|}{0.325} & 0.121 \\ \hline
\textbf{9} & 10 & 0.9 & \multicolumn{1}{r|}{13.901} & \multicolumn{1}{r|}{24.658} & \multicolumn{1}{r|}{0.003} & 0.002 & \multicolumn{1}{r|}{0.371} & 0.041 \\ \hline
\textbf{10} & 20 & 0.1 & \multicolumn{1}{r|}{0.006} & \multicolumn{1}{r|}{0.001} & \multicolumn{1}{r|}{0.002} & 0.000 & \multicolumn{1}{r|}{0.254} & 0.005 \\ \hline
\textbf{11} & 20 & 0.2 & \multicolumn{1}{r|}{0.034} & \multicolumn{1}{r|}{0.029} & \multicolumn{1}{r|}{0.003} & 0.001 & \multicolumn{1}{r|}{0.296} & 0.038 \\ \hline
\textbf{12} & 20 & 0.3 & \multicolumn{1}{r|}{0.192} & \multicolumn{1}{r|}{0.306} & \multicolumn{1}{r|}{0.006} & 0.001 & \multicolumn{1}{r|}{0.346} & 0.026 \\ \hline
\textbf{13} & 20 & 0.4 & \multicolumn{1}{r|}{0.875} & \multicolumn{1}{r|}{0.390} & \multicolumn{1}{r|}{0.012} & 0.007 & \multicolumn{1}{r|}{0.440} & 0.015 \\ \hline
\textbf{14} & 20 & 0.5 & \multicolumn{1}{r|}{10.721} & \multicolumn{1}{r|}{12.125} & \multicolumn{1}{r|}{0.019} & 0.013 & \multicolumn{1}{r|}{0.498} & 0.047 \\ \hline
\textbf{15} & 20 & 0.6 & \multicolumn{1}{r|}{124.385} & \multicolumn{1}{r|}{91.853} & \multicolumn{1}{r|}{0.023} & 0.012 & \multicolumn{1}{r|}{0.591} & 0.034 \\ \hline
\textbf{16} & 20 & 0.7 & \multicolumn{1}{r|}{199.375} & \multicolumn{1}{r|}{115.200} & \multicolumn{1}{r|}{0.034} & 0.015 & \multicolumn{1}{r|}{0.667} & 0.060 \\ \hline
\textbf{17} & 20 & 0.8 & \multicolumn{1}{r|}{300.065} & \multicolumn{1}{r|}{0.010} & \multicolumn{1}{r|}{0.034} & 0.018 & \multicolumn{1}{r|}{0.840} & 0.061 \\ \hline
\textbf{18} & 20 & 0.9 & \multicolumn{1}{r|}{300.094} & \multicolumn{1}{r|}{0.015} & \multicolumn{1}{r|}{0.040} & 0.014 & \multicolumn{1}{r|}{1.056} & 0.063 \\ \hline
\textbf{19} & 30 & 0.1 & \multicolumn{1}{r|}{0.019} & \multicolumn{1}{r|}{0.008} & \multicolumn{1}{r|}{0.006} & 0.003 & \multicolumn{1}{r|}{1.720} & 0.035 \\ \hline
\textbf{20} & 30 & 0.2 & \multicolumn{1}{r|}{0.340} & \multicolumn{1}{r|}{0.282} & \multicolumn{1}{r|}{0.015} & 0.010 & \multicolumn{1}{r|}{2.427} & 0.036 \\ \hline
\textbf{21} & 30 & 0.3 & \multicolumn{1}{r|}{9.814} & \multicolumn{1}{r|}{4.160} & \multicolumn{1}{r|}{0.035} & 0.012 & \multicolumn{1}{r|}{3.181} & 0.054 \\ \hline
\textbf{22} & 30 & 0.4 & \multicolumn{1}{r|}{267.903} & \multicolumn{1}{r|}{70.468} & \multicolumn{1}{r|}{0.070} & 0.024 & \multicolumn{1}{r|}{3.985} & 0.049 \\ \hline
\textbf{23} & 30 & 0.5 & \multicolumn{1}{r|}{300.103} & \multicolumn{1}{r|}{0.012} & \multicolumn{1}{r|}{0.138} & 0.083 & \multicolumn{1}{r|}{5.017} & 0.331 \\ \hline
\textbf{24} & 30 & 0.6 & \multicolumn{1}{r|}{300.186} & \multicolumn{1}{r|}{0.038} & \multicolumn{1}{r|}{0.163} & 0.040 & \multicolumn{1}{r|}{6.124} & 0.287 \\ \hline
\textbf{25} & 30 & 0.7 & \multicolumn{1}{r|}{300.254} & \multicolumn{1}{r|}{0.042} & \multicolumn{1}{r|}{0.173} & 0.054 & \multicolumn{1}{r|}{7.397} & 0.534 \\ \hline
\textbf{26} & 30 & 0.8 & \multicolumn{1}{r|}{300.403} & \multicolumn{1}{r|}{0.057} & \multicolumn{1}{r|}{0.189} & 0.077 & \multicolumn{1}{r|}{9.811} & 0.697 \\ \hline
\textbf{27} & 30 & 0.9 & \multicolumn{1}{r|}{300.556} & \multicolumn{1}{r|}{0.122} & \multicolumn{1}{r|}{0.304} & 0.111 & \multicolumn{1}{r|}{12.687} & 1.312 \\ \hline
\textbf{28} & 40 & 0.1 & \multicolumn{1}{c|}{-} & - & \multicolumn{1}{r|}{0.021} & 0.009 & \multicolumn{1}{r|}{2.366} & 0.267 \\ \hline
\textbf{29} & 40 & 0.2 & \multicolumn{1}{c|}{-} & - & \multicolumn{1}{r|}{0.067} & 0.038 & \multicolumn{1}{r|}{3.753} & 0.340 \\ \hline
\textbf{30} & 40 & 0.3 & \multicolumn{1}{c|}{-} & - & \multicolumn{1}{r|}{0.191} & 0.104 & \multicolumn{1}{r|}{5.472} & 0.367 \\ \hline
\textbf{31} & 40 & 0.4 & \multicolumn{1}{c|}{-} & - & \multicolumn{1}{r|}{0.753} & 0.932 & \multicolumn{1}{r|}{6.545} & 0.542 \\ \hline
\textbf{32} & 40 & 0.5 & \multicolumn{1}{c|}{-} & - & \multicolumn{1}{r|}{\textit{\textbf{0.867}}} & 0.631 & \multicolumn{1}{r|}{8.526} & 0.695 \\ \hline
\textbf{33} & 40 & 0.6 & \multicolumn{1}{c|}{-} & - & \multicolumn{1}{r|}{1.306} & 1.014 & \multicolumn{1}{r|}{10.350} & 0.448 \\ \hline
\textbf{34} & 40 & 0.7 & \multicolumn{1}{c|}{-} & - & \multicolumn{1}{r|}{1.557} & 2.098 & \multicolumn{1}{r|}{13.801} & 2.007 \\ \hline
\textbf{35} & 40 & 0.8 & \multicolumn{1}{c|}{-} & - & \multicolumn{1}{r|}{\textit{\textbf{1.162}}} & 0.431 & \multicolumn{1}{r|}{17.844} & 1.989 \\ \hline
\textbf{36} & 40 & 0.9 & \multicolumn{1}{c|}{-} & - & \multicolumn{1}{r|}{1.155} & 0.159 & \multicolumn{1}{r|}{24.713} & 3.280 \\ \hline
\textbf{37} & 50 & 0.1 & \multicolumn{1}{c|}{-} & - & \multicolumn{1}{r|}{0.027} & 0.014 & \multicolumn{1}{r|}{3.607} & 0.053 \\ \hline
\textbf{38} & 50 & 0.2 & \multicolumn{1}{c|}{-} & - & \multicolumn{1}{r|}{0.333} & 0.245 & \multicolumn{1}{r|}{5.174} & 0.192 \\ \hline
\textbf{39} & 50 & 0.3 & \multicolumn{1}{c|}{-} & - & \multicolumn{1}{r|}{4.990} & 5.810 & \multicolumn{1}{r|}{7.463} & 0.453 \\ \hline
\textbf{40} & 50 & 0.4 & \multicolumn{1}{c|}{-} & - & \multicolumn{1}{r|}{4.757} & 4.187 & \multicolumn{1}{r|}{9.495} & 0.510 \\ \hline
\textbf{41} & 50 & 0.5 & \multicolumn{1}{c|}{-} & - & \multicolumn{1}{r|}{7.071} & 4.687 & \multicolumn{1}{r|}{14.036} & 1.771 \\ \hline
\textbf{42} & 50 & 0.6 & \multicolumn{1}{c|}{-} & - & \multicolumn{1}{r|}{66.279} & 61.284 & \multicolumn{1}{r|}{25.216} & 5.909 \\ \hline
\textbf{43} & 50 & 0.7 & \multicolumn{1}{c|}{-} & - & \multicolumn{1}{r|}{38.152} & 46.330 & \multicolumn{1}{r|}{25.096} & 3.382 \\ \hline
\textbf{44} & 50 & 0.8 & \multicolumn{1}{c|}{-} & - & \multicolumn{1}{r|}{\textit{\textbf{7.727}}} & 8.428 & \multicolumn{1}{r|}{33.786} & 8.187 \\ \hline
\textbf{45} & 50 & 0.9 & \multicolumn{1}{c|}{-} & - & \multicolumn{1}{r|}{2.372} & 0.257 & \multicolumn{1}{r|}{42.495} & 4.156 \\ \hline
\textbf{46} & 60 & 0.1 & \multicolumn{1}{c|}{-} & - & \multicolumn{1}{r|}{0.129} & 0.087 & \multicolumn{1}{r|}{4.421} & 0.129 \\ \hline
\textbf{47} & 60 & 0.2 & \multicolumn{1}{c|}{-} & - & \multicolumn{1}{r|}{\textit{\textbf{0.872}}} & 1.026 & \multicolumn{1}{r|}{7.734} & 1.722 \\ \hline
\textbf{48} & 60 & 0.3 & \multicolumn{1}{c|}{-} & - & \multicolumn{1}{r|}{16.930} & 11.034 & \multicolumn{1}{r|}{11.445} & 1.823 \\ \hline
\textbf{49} & 60 & 0.4 & \multicolumn{1}{c|}{-} & - & \multicolumn{1}{r|}{51.511} & 82.086 & \multicolumn{1}{r|}{18.106} & 4.262 \\ \hline
\textbf{50} & 60 & 0.5 & \multicolumn{1}{c|}{-} & - & \multicolumn{1}{r|}{68.744} & 68.795 & \multicolumn{1}{r|}{21.368} & 3.984 \\ \hline
\textbf{51} & 60 & 0.6 & \multicolumn{1}{c|}{-} & - & \multicolumn{1}{r|}{95.411} & 88.234 & \multicolumn{1}{r|}{48.351} & 13.262 \\ \hline
\textbf{52} & 60 & 0.7 & \multicolumn{1}{c|}{-} & - & \multicolumn{1}{r|}{66.161} & 44.420 & \multicolumn{1}{r|}{40.763} & 11.270 \\ \hline
\textbf{53} & 60 & 0.8 & \multicolumn{1}{c|}{-} & - & \multicolumn{1}{r|}{42.528} & 68.171 & \multicolumn{1}{r|}{56.013} & 10.085 \\ \hline
\textbf{54} & 60 & 0.9 & \multicolumn{1}{c|}{-} & - & \multicolumn{1}{r|}{\textit{\textbf{7.724}}} & 1.407 & \multicolumn{1}{r|}{72.302} & 7.284 \\ \hline
\textbf{55} & 70 & 0.1 & \multicolumn{1}{c|}{-} & - & \multicolumn{1}{r|}{0.461} & 0.291 & \multicolumn{1}{r|}{6.458} & 1.655 \\ \hline
\textbf{56} & 70 & 0.2 & \multicolumn{1}{c|}{-} & - & \multicolumn{1}{r|}{4.822} & 4.982 & \multicolumn{1}{r|}{9.977} & 1.664 \\ \hline
\textbf{57} & 70 & 0.3 & \multicolumn{1}{c|}{-} & - & \multicolumn{1}{r|}{28.828} & 14.424 & \multicolumn{1}{r|}{\textit{\textbf{15.687}}} & 2.101 \\ \hline
\textbf{58} & 70 & 0.4 & \multicolumn{1}{c|}{-} & - & \multicolumn{1}{r|}{78.486} & 63.512 & \multicolumn{1}{r|}{38.800} & 14.893 \\ \hline
\textbf{59} & 70 & 0.5 & \multicolumn{1}{c|}{-} & - & \multicolumn{1}{r|}{114.880} & 70.768 & \multicolumn{1}{r|}{38.288} & 14.127 \\ \hline
\textbf{60} & 70 & 0.6 & \multicolumn{1}{c|}{-} & - & \multicolumn{1}{r|}{103.728} & 71.210 & \multicolumn{1}{r|}{\textit{\textbf{54.067}}} & 11.507 \\ \hline
\textbf{61} & 70 & 0.7 & \multicolumn{1}{c|}{-} & - & \multicolumn{1}{r|}{133.313} & 72.464 & \multicolumn{1}{r|}{63.095} & 10.184 \\ \hline
\textbf{62} & 70 & 0.8 & \multicolumn{1}{c|}{-} & - & \multicolumn{1}{r|}{110.066} & 77.881 & \multicolumn{1}{r|}{96.272} & 17.250 \\ \hline
\textbf{63} & 70 & 0.9 & \multicolumn{1}{c|}{-} & - & \multicolumn{1}{r|}{\textit{\textbf{19.536}}} & 4.820 & \multicolumn{1}{r|}{134.355} & 28.658 \\ \hline
\textbf{64} & 80 & 0.1 & \multicolumn{1}{c|}{-} & - & \multicolumn{1}{r|}{\textit{\textbf{1.309}}} & 1.599 & \multicolumn{1}{r|}{7.324} & 0.592 \\ \hline
\textbf{65} & 80 & 0.2 & \multicolumn{1}{c|}{-} & - & \multicolumn{1}{r|}{52.218} & 78.163 & \multicolumn{1}{r|}{15.138} & 2.710 \\ \hline
\textbf{66} & 80 & 0.3 & \multicolumn{1}{c|}{-} & - & \multicolumn{1}{r|}{67.343} & 36.356 & \multicolumn{1}{r|}{26.198} & 6.943 \\ \hline
\textbf{67} & 80 & 0.4 & \multicolumn{1}{c|}{-} & - & \multicolumn{1}{r|}{116.747} & 111.805 & \multicolumn{1}{r|}{\textit{\textbf{37.085}}} & 9.220 \\ \hline
\textbf{68} & 80 & 0.5 & \multicolumn{1}{c|}{-} & - & \multicolumn{1}{r|}{96.057} & 37.923 & \multicolumn{1}{r|}{\textit{\textbf{49.570}}} & 7.162 \\ \hline
\textbf{69} & 80 & 0.6 & \multicolumn{1}{c|}{-} & - & \multicolumn{1}{r|}{167.210} & 66.138 & \multicolumn{1}{r|}{159.555} & 81.808 \\ \hline
\textbf{70} & 80 & 0.7 & \multicolumn{1}{c|}{-} & - & \multicolumn{1}{r|}{202.061} & 65.897 & \multicolumn{1}{r|}{112.915} & 18.953 \\ \hline
\textbf{71} & 80 & 0.8 & \multicolumn{1}{c|}{-} & - & \multicolumn{1}{r|}{\textit{\textbf{162.988}}} & 45.676 & \multicolumn{1}{r|}{167.062} & 37.764 \\ \hline
\textbf{72} & 80 & 0.9 & \multicolumn{1}{c|}{-} & - & \multicolumn{1}{r|}{\textit{\textbf{53.323}}} & 18.095 & \multicolumn{1}{r|}{227.224} & 50.657 \\ \hline
\textbf{73} & 90 & 0.1 & \multicolumn{1}{c|}{-} & - & \multicolumn{1}{r|}{1.058} & 1.224 & \multicolumn{1}{r|}{9.336} & 0.726 \\ \hline
\textbf{74} & 90 & 0.2 & \multicolumn{1}{c|}{-} & - & \multicolumn{1}{r|}{34.024} & 38.601 & \multicolumn{1}{r|}{18.623} & 3.323 \\ \hline
\textbf{75} & 90 & 0.3 & \multicolumn{1}{c|}{-} & - & \multicolumn{1}{r|}{162.527} & 97.995 & \multicolumn{1}{r|}{\textit{\textbf{32.050}}} & 9.461 \\ \hline
\textbf{76} & 90 & 0.4 & \multicolumn{1}{c|}{-} & - & \multicolumn{1}{r|}{143.790} & 114.855 & \multicolumn{1}{r|}{\textit{\textbf{48.503}}} & 4.782 \\ \hline
\textbf{77} & 90 & 0.5 & \multicolumn{1}{c|}{-} & - & \multicolumn{1}{r|}{167.725} & 63.551 & \multicolumn{1}{r|}{\textit{\textbf{67.405}}} & 14.039 \\ \hline
\textbf{78} & 90 & 0.6 & \multicolumn{1}{c|}{-} & - & \multicolumn{1}{r|}{157.333} & 100.262 & \multicolumn{1}{r|}{\textit{\textbf{81.155}}} & 17.398 \\ \hline
\textbf{79} & 90 & 0.7 & \multicolumn{1}{c|}{-} & - & \multicolumn{1}{r|}{252.613} & 37.275 & \multicolumn{1}{r|}{\textit{\textbf{90.603}}} & 16.960 \\ \hline
\textbf{80} & 90 & 0.8 & \multicolumn{1}{c|}{-} & - & \multicolumn{1}{r|}{258.383} & 18.204 & \multicolumn{1}{r|}{141.974} & 29.896 \\ \hline
\textbf{81} & 90 & 0.9 & \multicolumn{1}{c|}{-} & - & \multicolumn{1}{r|}{\textit{\textbf{120.630}}} & 47.417 & \multicolumn{1}{r|}{219.341} & 43.442 \\ \hline
\textbf{82} & 100 & 0.1 & \multicolumn{1}{c|}{-} & - & \multicolumn{1}{r|}{1.571} & 0.714 & \multicolumn{1}{r|}{11.952} & 2.225 \\ \hline
\textbf{83} & 100 & 0.2 & \multicolumn{1}{c|}{-} & - & \multicolumn{1}{r|}{90.416} & 70.264 & \multicolumn{1}{r|}{23.365} & 2.986 \\ \hline
\textbf{84} & 100 & 0.3 & \multicolumn{1}{c|}{-} & - & \multicolumn{1}{r|}{106.807} & 100.061 & \multicolumn{1}{r|}{\textit{\textbf{48.057}}} & 10.612 \\ \hline
\textbf{85} & 100 & 0.4 & \multicolumn{1}{c|}{-} & - & \multicolumn{1}{r|}{127.184} & 107.183 & \multicolumn{1}{r|}{\textit{\textbf{60.967}}} & 10.674 \\ \hline
\textbf{86} & 100 & 0.5 & \multicolumn{1}{c|}{-} & - & \multicolumn{1}{r|}{92.338} & 108.744 & \multicolumn{1}{r|}{112.470} & 12.526 \\ \hline
\textbf{87} & 100 & 0.6 & \multicolumn{1}{c|}{-} & - & \multicolumn{1}{r|}{171.188} & 79.565 & \multicolumn{1}{r|}{\textit{\textbf{105.843}}} & 17.305 \\ \hline
\textbf{88} & 100 & 0.7 & \multicolumn{1}{c|}{-} & - & \multicolumn{1}{r|}{138.741} & 106.281 & \multicolumn{1}{r|}{155.752} & 41.929 \\ \hline
\textbf{89} & 100 & 0.8 & \multicolumn{1}{c|}{-} & - & \multicolumn{1}{r|}{184.035} & 128.311 & \multicolumn{1}{r|}{214.694} & 59.357 \\ \hline
\textbf{90} & 100 & 0.9 & \multicolumn{1}{c|}{-} & - & \multicolumn{1}{r|}{\textit{\textbf{233.650}}} & 54.277 & \multicolumn{1}{r|}{281.240} & 33.966 \\ \hline
\end{longtable}

\end{widetext}

\end{document}